\documentclass[a4paper, 5p]{elsarticle}
\usepackage{epsfig}
\usepackage{amsmath}
\usepackage{hyperref}
\usepackage{graphicx}

\journal{Physica A}

\begin{document}

\begin{frontmatter}

\title{The Origin of Power-Law Emergent Scaling in Large Binary Networks}

\author[cnm]{D.~P.~Almond}
\author[cnm]{C.~J.~Budd}
\author[cnm]{M.~A. Freitag}
\author[cnm]{G.~W.~Hunt}
\author[cnm,leeds]{N.~J.~McCullen\corref{cor1}}
\ead{n.mccullen@physics.org}
\author[cnm]{N.~D.~Smith}

\address[cnm]{Center for Nonlinear Mechanics, University of Bath, BA2 7AY, UK}
\address[leeds]{School of Mathematics, University of Leeds, LS2 9JT, UK}

\cortext[cor1]{Corresponding author}

\begin{abstract}
In this paper we study the macroscopic conduction properties of large but finite binary networks with conducting bonds. 
By taking a combination of a spectral and an averaging based approach we derive asymptotic formulae for the conduction in terms of the component proportions $p$ and the total number of components $N$. 
These formul\ae\ correctly identify both the percolation limits and also the emergent power law behaviour between the percolation limits and show the interplay between the size of the network and the deviation of the proportion from the critical value of $p = 1/2$. 
The results compare excellently with a large number of numerical simulations.
\end{abstract}


\end{frontmatter}

\section{Introduction and summary}
 
Large but finite binary networks comprising disordered mixtures of two interacting components can arise both directly, in electrical circuits \cite{bouamrane2003esp,clerc1990ecb,vainas1999err,Kirk} or mechanical structures \cite{murphy2006ees}, and as models of other systems such as disordered materials with varying electrical \cite{dyre2000uac}, thermal, mechanical or even geophysical properties in the micro-scale, coupled at a meso-scale \cite{Jurga}.
They are prototypes of many forms of {\em complex systems} which  are often observed to have {\em macroscopic emergent properties} which can have {\em emergent power-law behaviour} over a wide range of parameter values which is different from any power law behaviour of the individual elements of the network, and is a consequence of the way in which the responses of the components {\em combine}. 
For certain ranges of parameters we see the extensively studied {\em percolation type} of behaviour \cite{Grim}, in which the overall conductance is directly proportional to the individual component conductances with a constant of proportionality dependent both on the component proportion and on the network size. 
As well as being important in their own right, such large binary networks also provide a useful test bed for identifying different types of emergent behaviour, determining what causes it, finding the range of parameters over which it applies and addressing the fundamental question of which aspects of a complex system, such as the number and proportion of the components, lead to, and influence, the emergent behaviour. 
In this paper, we will combine a spectral analysis, motivated by \cite{jonckheere1998drb}), of the (partly random) linear operators (Kirchhoff-type matrices) associated with the network, with the averaging methods described in \cite{Kirk}, to derive an asymptotic formula for the emergent network admittance, that includes both the effects of the component proportion $p$ and the network size $N$. 

\begin{figure}
\centering
\includegraphics[width=0.8\linewidth]{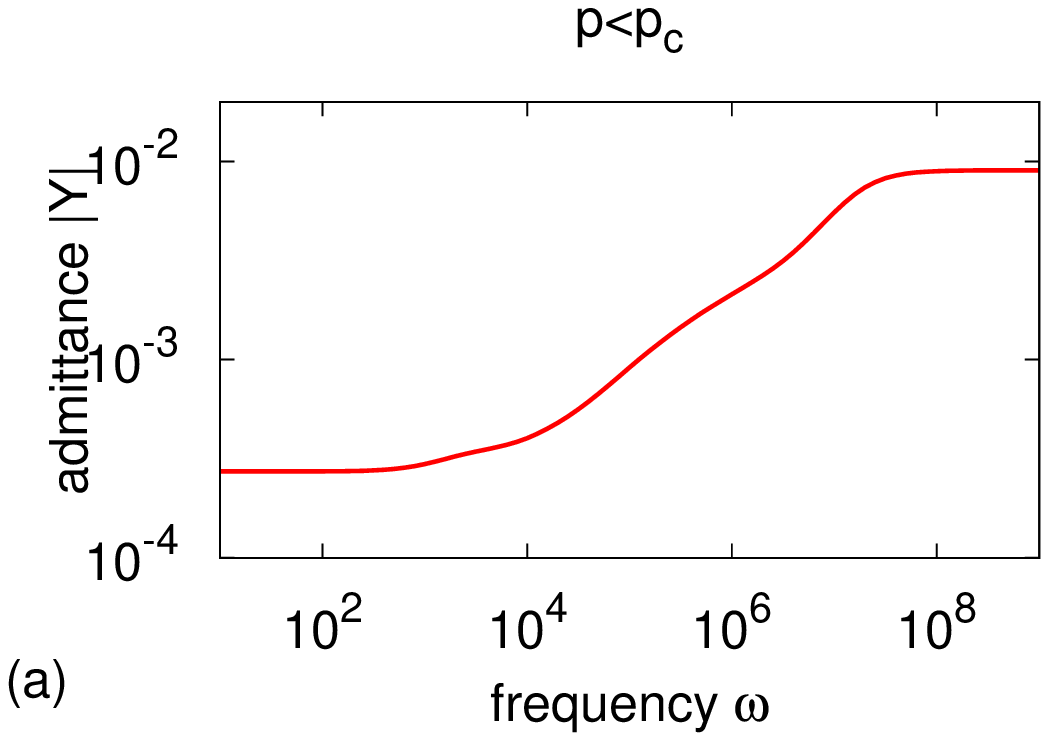}
\includegraphics[width=0.8\linewidth]{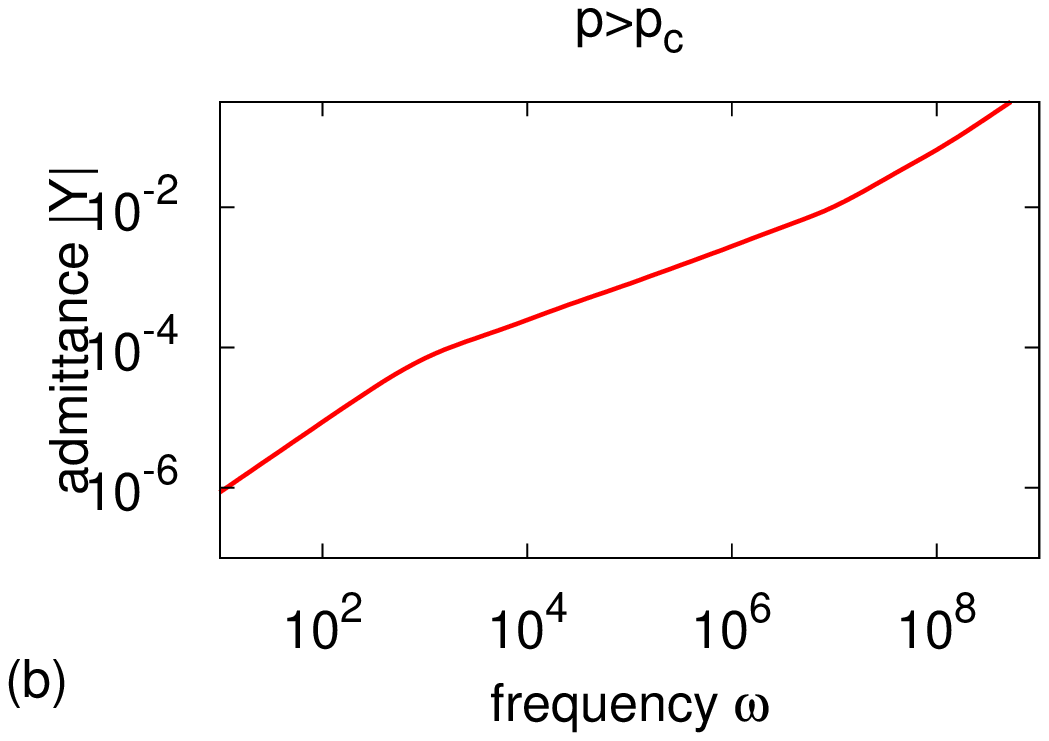}
\caption{(color online) Examples of the type of response observed for the proportion $p$ of capacitors (a) below and (b) above the percolation threshold $p = p_c = 1/2$.}\label{fig:exfig}
\end{figure}

As an example, we  consider (a set of random realisations of) a binary square network comprising a random mixture of $N$ conducting bonds which are either chosen to have a constant conductance $y_1 = 1/R$ or a variable complex admittance $y_2 = i \omega C$, which is directly proportional to an angular frequency $\omega$. 
If $p$ is the occupation probability for choosing a $y_2$ component (and $(1-p)$ the occupation probability for $y_1$), in the limit of large $N$ or for averages over large numbers of systems, it directly determines the proportion of $y_2$ to be approximately $p$ (and $y_1$ to be $(1-p))$.
This network, when subjected to  an applied alternating voltage of angular frequency $\omega$, has macroscopic features, such as the total admittance $Y(\omega)$.  Over a wide range of frequencies $0 < \omega_1 < \omega < \omega_2$, the admittance displays {\em power law emergent characteristics} so that the magnitude of the complex admittance $|Y|$ is proportional to $\omega^{\alpha}$ for some $\alpha(p)$.
The effects of network size, and component proportion, are important in that $\omega_1$ and $\omega_2$ depend upon both $p$ and $N$ and it is well known \cite{Grim} that the case $p=1/2$ is a critical value ($p_c$) for two-dimensional square networks.
If $p \ne 1/2$ and $N$ is sufficiently large then this problem can be studied by averaging \cite{Kirk}, with $\omega_1 \to 0$ and $\omega_2 \to \infty$ as $p \to 1/2$.
In contrast, if $p = 1/2$, $\omega_1$ is inversely proportional to $N$ and $\omega_2$ directly proportional to $N$, as $N$ increases to infinity. 
For $0 < \omega < \omega_1$ and $\omega > \omega_2$ percolation type behaviour is observed  for which $Y$ is proportional to either $y_1$ or $y_2$ with a constant of proportionality subtly dependent on $|p-1/2|$.
Hence we see in this system (i) an emergent region with a power law response depending on the proportion but not the arrangement or number of the components (ii) a more random region (iii) a transition between these two regions at frequency values which depends  on the number and proportion of components in the system. An illustration of the different types of observed response is shown in Figure \ref{fig:exfig}.

The purpose of this paper is to give insight into this behaviour by obtaining asymptotic formul\ae\ for the expected response curves.
We compare and extend results obtained by two complementary methods, one based upon averaging \cite{Kirk} and the other based on  properties of the spectrum of certain operators \cite{jonckheere1998drb}.  
The averaging method works well when $p \ne 1/2$ and $N \to \infty$, and the spectral method, in contrast, works well for  the case of $ p = 1/2$ and large, finite $N$. 
By combining the averaging method with the spectral results we will also give approximate asymptotic formul\ae\, (\ref{rainy1a},\ref{rainy2a}) for $Y$,  valid for general values of $p \approx 1/2$ and sufficiently large $N$, which demonstrates the interplay between network size and component proportion.
The spectral method is based both on rigorous results concerning the poles and zero distribution of the function $Y(\omega)$ and on certain semi-empirical results on the  regularity of their statistical distribution.
From these observations we derive, in the case of $p = p_c = 1/2$ an asymptotic form for the response which gives results  almost indistinguishable from the numerical simulations,
showing that  a power law response in which $|Y(\omega)|$ is equal to $\sqrt{\omega C/R}$, is observed over a range 
$$1/NCR = \omega_1 < \omega < \omega_2 = N/CR.$$ 
For $\omega CR > N$ or $\omega CR < 1/N$ this is replaced by a `percolation' type response for which $|Y|$ is proportional to one of
\begin{align}
1/(\sqrt{N}R)\,\, &\mbox{or}\,\,\sqrt{N} \omega C \quad \mbox{for} \quad \omega \ll 1,\nonumber\\
&\mbox{and}&\nonumber\\
\sqrt{N}/R\,\,&\mbox{or}\,\,\omega C/\sqrt{N}  \quad \mbox{for} \quad \omega \gg 1.
\label{snow0}
\end{align}

When $p \ne 1/2$ we find, both numerically and asymptotically, that for $p \approx 1/2$ very similar results are obtained  to those obtained for $p = 1/2$ if $1 \ll N < |1/2 - p|^{-2}$.
For larger values of $N$ the response becomes independent of $N$ and asymptotic to a  $p$ dependent response. For values of $p$
not too close to 1/2 this behaviour is approximately predicted by the Effective-Medium-Approximation (EMA)  homogenisation approach based on averaging \cite{milton1980bcd}, \cite{Kirk}.
For large and small values of $\omega$, the EMA predicts percolation limits in which $|Y|$ is proportional to 
\begin{align}
(1-2p)/ R  \quad &\mbox{for} \quad \omega \ll \omega_1,\nonumber\\
&\mbox{and}&\nonumber\\
\quad 1/((1-2p)R)\quad &\mbox{for} \quad \omega \gg \omega_2\nonumber\\
&\mbox{if}\quad p < 1/2,
\label{snow1}
\end{align}
and
\begin{align}
\omega C/(2p-1)  \quad &\mbox{for} \quad \omega \ll \omega_1,\nonumber\\
&\mbox{and}&\nonumber\\
 \quad (2p-1)\omega C   \quad &\mbox{for} \quad \omega \gg \omega_2 \nonumber\\
  &\mbox{if} \quad p > 1/2.
\label{snow2}
\end{align}
The percolation limits $\omega_1(p) < \omega_2(p)$ satisfy $\omega_1(p) \to 0, \omega_2(p) \to \infty$ as $p \to 1/2$.
In the power law emergent region $\omega_1 < \omega < \omega_2$ we see power law behaviour proportional to $\omega^{\alpha}$. 
For C-R networks we will show that the EMA  implies that
$$\alpha = \frac{1}{2} - \frac{\epsilon}{2 \sqrt{2} \sqrt{1 - \epsilon^2/2}} \approx p, \quad \epsilon = 1 - 2p$$
and for R-R networks with $y_2 = \mu y_1$ with $\mu$ real, we see power law behaviour proportional to  $\mu^p$.
As described above, we will consider a combination of the EMA and spectral approaches which allows for finite size effects and gives formul\ae\, (\ref{rainy1a},\ref{rainy2a}) for $Y$ involving $N$ and $p$,
in which both (\ref{snow0},\ref{snow1},\ref{snow2}) arise as
special cases. 
We note, however, that the EMA prediction is not particularly good in the limit of $p \to 1/2$. In particular, it has been observed empirically \cite{jonckheere1998drb}, that rather than having percolation limits
proportional to $|1/2 - p|$ or $|1/2 - p|^{-1}$, in the limit of $|1/2 - p| \ll 1$ they are more closely approximated by expressions of the form
\begin{equation}
{
|Y| \sim |1/2 - p|^{\pm \beta}, \quad \beta \approx 1.3. 
}
\label{1.3eqn}
\end{equation}

The layout of the remainder of this paper is as follows.
In Section \ref{sec:models} we will give a series of numerical results for a general binary network with admittances $y_1$ and $y_2$, 
which illustrate the various points made above on the nature of the network response and will look at both power law emergent behaviour and
at percolation responses.
In Section \ref{sec:lincirc} we will formulate the matrix equations describing the network and will show how the poles and zeros of the admittance function interlace. In Section \ref{sec:distributions} we will discuss, and derive, a series of statistical results concerning the distribution of the poles ad zeros. 
In Section \ref{sec:asymp} we will use these statistical results to derive a precise  asymptotic form of the admittance $Y$ of a general binary network, when $p = 1/2$ and $N$ is large. In Section \ref{sec:ema} we review the (classical) averaging method for $N = \infty$ which gives an excellent estimate when $p$ is not too close to $1/2$, and will also consider a combination of  this method with the spectral method for finite $N$, leading
to the formul\ae\ (\ref{rainy1a},\ref{rainy2a}) for the response for all $p$ and sufficiently large $N$. In Section \ref{sec:comparison} we compare the predictions of the asymptotic formul\ae\ with numerical computations of the network responses.
Finally in Section \ref{sec:concs} we will draw some conclusions from this work.

\section{Network models and their responses}\label{sec:models}

In this section we consider basic models for composite materials and associated random binary electrical networks with bonds having admittance $y_1$
and $y_2$, and present the graphs of their responses. 
In particular we will look in detail at the existence of a power law emergent region, and will obtain empirical evidence for the effects of network size $N$ and capacitor  proportion $p$, on  both this region and the `percolation behaviour' when 
$|\mu| = |y_2/y_1|$ is either large or small.

\subsection{Composite materials and their properties}

An initial motivation for studying binary networks comes from models of composite materials. 
Disordered two-phase composite materials are found to exhibit power-law scaling in their bulk responses over several orders of magnitude in the contrast ratio of the components \cite{jonscher1996url, clerc1990ecb}, and this effect has been observed \cite{bouamrane2003esp, brosseau2005mas} in both physical and numerical experiment son conductor-dielectric composite materials. 
In the electrical experiments this was previously referred to as ``Universal Dielectric Response'' (UDR), and it has been observed \cite{ngai1979oud, jonscher1977udr} that this is an emergent property arising out of  the random nature of the mixture.
The same response is also found in numerical studies using 2D lattice structures with bonds randomly assigned to have a conductivity which is either constant or linearly variable in a parameter.
These studies reveal that the emergent scaling is a property of a large number of complex system that can be represented as such a binary percolation network.
A simple model of such conductor-dielectric mixtures with fine structure is a large electrical circuit representation, replacing the constituent conducting and dielectric parts with a linear C-R network with $N \gg 1$ resistors and capacitors, respectively
forming the bonds in this network.
Similarly, large mixtures of materials with different resistance can be modelled by R-R resistor-resistor networks.
For a binary disordered mixture, the different components can then be assigned randomly to the bonds on the lattice \cite{truong1995ccc}.
In most previous studies a 2D square lattice has been used, with bonds assigned randomly as either C or R, with probability $p$, $1-p$ respectively.
The components are distributed in a two-dimensional lattice between two bus-bars. On of which is grounded and the other is raised to a potential $V(t) = V \exp(i\omega t)$.
This leads to a current $I(t) = I(\omega) \exp(i \omega t )$ between the bus-bars,  and we measure the macroscopic (complex) admittance given by 
$$Y(\omega) = I(\omega)/V.$$
This approach is closely related to percolation models and a large review of this and binary disordered networks can be found in \cite{jonckheere1998drb,clerc1990ecb,Grim}.
There are many advantages to using C-R network representations of these types of system.
In particular, widely available circuit simulation software can be used, which makes use of the available efficient sparse-matrix techniques in solving the equations of the system.
Additionally, algorithms based on the Frank-Lobb reduction techniques \cite{frank1988hea} can be used on the 2D square lattice representations to solve large systems efficiently \cite{bouamrane2003esp}.
These techniques were used in various studies to show that the PLER exists in any binary random network with variable contrast ratio \cite{truong1995ccc, almond1999dpr, vainas1999err, murphy2006ees, almond2006cdc}.
This allows many different simulations to be made of different realisations of the circuit with randomly assigned resistors and capacitors. 
Furthermore, finite element calculations reported in \cite{almond2006cdc} indicate that the response of the full material is very close to that of the network model of that material.

\subsection{Percolation and power-law emergent behaviour}

When  $\omega \ll 1$, the capacitors act as open circuits and conduction occurs predominantly through the resistors, having far higher admittance than the capacitors.
The circuit then becomes a {\em percolation network} in which the bonds are either conducting with probability $(1-p)$ or non-conducting with probability $p$.
The network only conducts only if there is a percolation path from one electrode to the other, and it is well known \cite{broadbent1953ppi} that, for 2D square lattices the critical percolation probability $p_c = 1/2$. 
Thus, in cases where the probability of the non-conducting phase $p > p_c = 1/2$, the conducting phase has a very low probability for such a percolation path to exist.

\begin{figure}
\centering
 \includegraphics[width=0.8\linewidth]{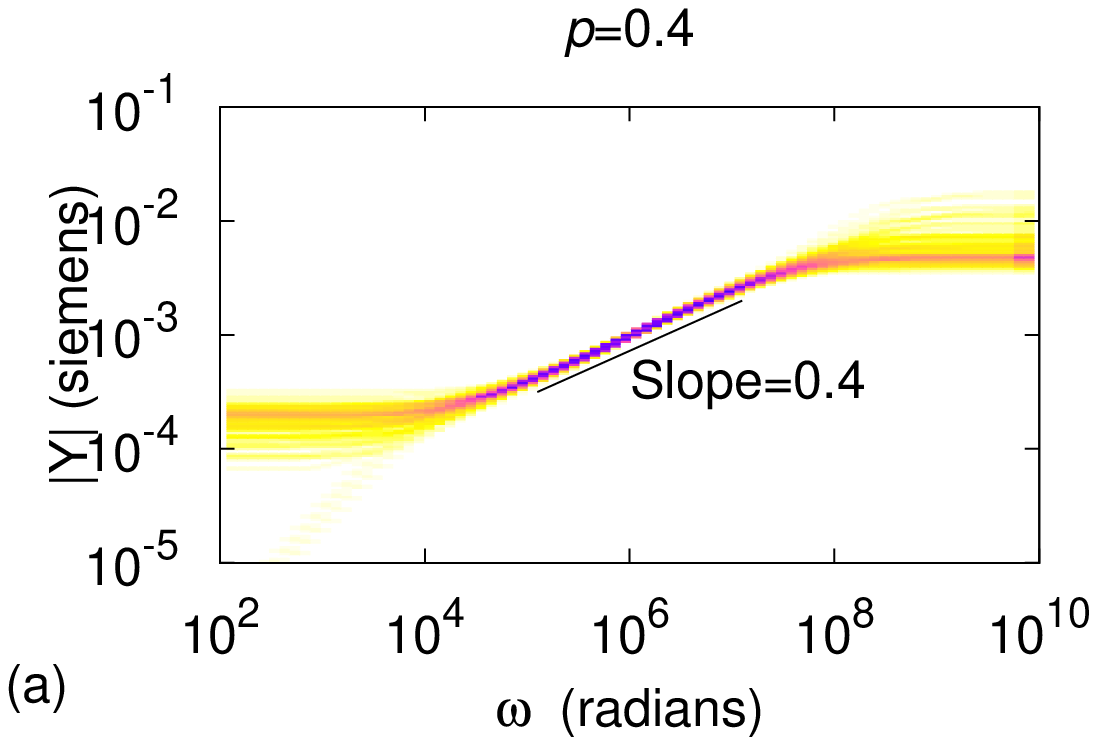}
 \includegraphics[width=0.8\linewidth]{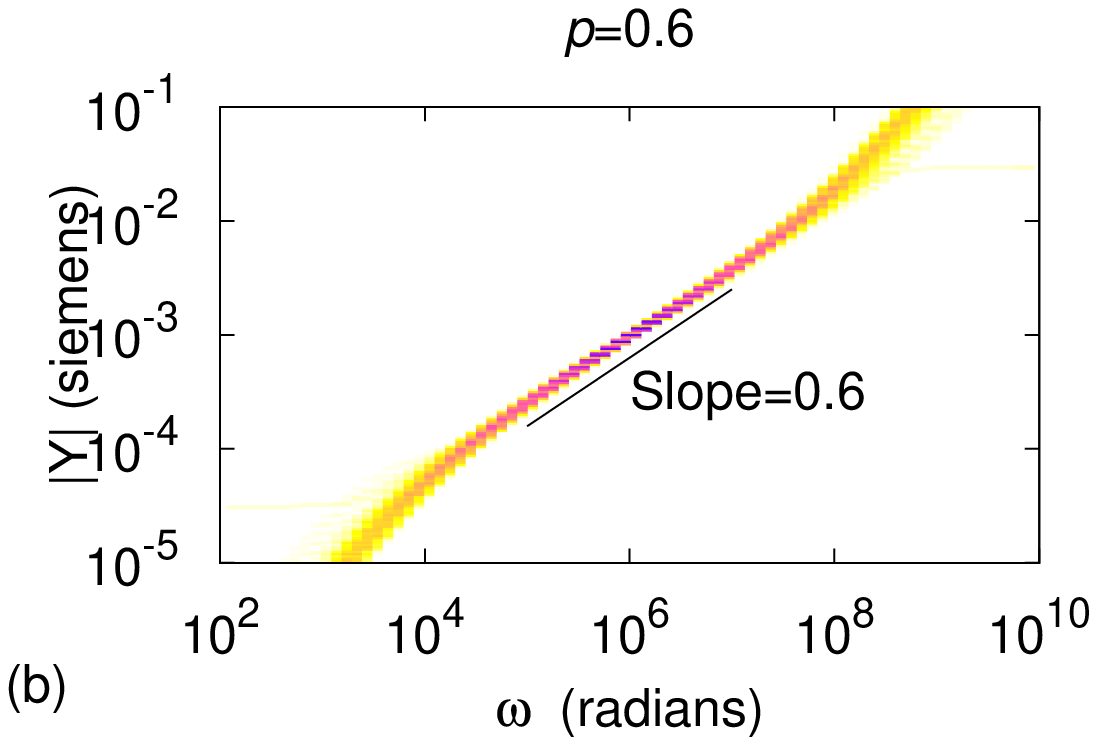}
\caption{(color online) Typical responses of network simulations for values of $p \ne 1/2$ which give qualitatively different behaviour so that in the percolation region with $\omega \ll 1$ or $\omega \gg 1$, we see resistive behaviour in case (a) and capacitative behaviour in case (b). 
The figures  presented are density plots of 100 random realisations for a $20\times 20$ network. 
Note that all of the realisations give very similar results.}
\label{fig:response46}
\end{figure}

\begin{figure}
\centering
 \includegraphics[width=0.8\linewidth]{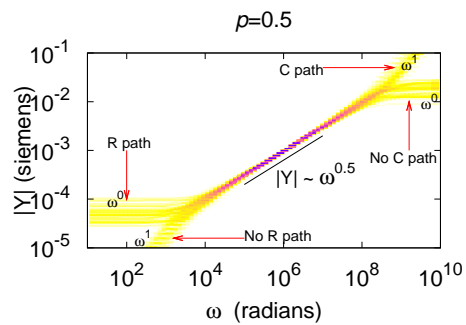}
\caption{(color online) Responses for 100 realisations at $p=1/2$ showing four different qualitative types of response for different realisations. 
Here, about half of the responses have a resistive percolation path and half have a capacitive one at low frequencies with a similar behaviour at high frequencies.  
The responses at high and low $\omega$ indicate which of these cases exist for a particular realisation.
The power-law emergent region can also be seen in which the admittance scales as $\sqrt{\omega}$ and all of the responses of the different network realisations coincide}
\label{fig:response5}
\end{figure}

In contrast, if $p < 1/2$ then such a path exists with probability approaching one as the network size increases. 
The case of $p = 1/2$ is critical with a 50\% probability that such a path exists. 
This implies that if $p < 1/2$ there is almost certainly only a resistive conduction path  and for low frequencies the overall admittance is independent of $\omega$.
In contrast, if $p > 1/2$ then there is almost certainly no path through the resistors and the admittance is capacitance dominated and directly proportional to $\omega$.
If $p = 1/2$ then half of the realisations will give an admittance response independent of $\omega$ and half an admittance response proportional to $\omega$. When $\omega \gg 1$, we see an opposite response.
In this case the capacitors act as almost short circuits with far higher admittance than the resistors. 
Again we see percolation behaviour with the resistors behaving approximately as open circuits in this case. 
Thus if $p > 1/2$ any conduction path is most likely solely capacitative, with the resulting overall admittance being proportional to $\omega$, and if $p < 1/2$ a response independent of $\omega$. 
The case of $p = 1/2$ again leads to both types of response having equal likelihood of occurrence, depending upon the network configuration. 
Note that this implies that if $p = 1/2$ then there are {\em four possible qualitatively different} types of response for any random realisation of the system.
For {\em intermediate values}  of $\omega$ the values of the admittance of the resistors and the capacitors are much closer to each other. 
It is here that we see power-law emergent behaviour (PLER).
This is characterised by two features: 
{\em (i)} an admittance response $|Y|$ that is proportional to $\omega^{\alpha}, \alpha \approx p$ over a range $\omega \in (\omega_1,\omega_2)$ and which displays a strong
symmetry in the behaviour for small and large values of $\omega$.
{\em (ii)} when $p = 1/2$ a response that is not randomly dependent upon the network configuration.
Figures \ref{fig:response46} (a) and (b) plot the admittance response as a function of $\omega$ in the cases of $p = 0.4$, $p = 0.6$ and Figure \ref{fig:response5} for $p = 1/2$. 
The figures clearly demonstrate the forms of behaviour described above. 
Observe that in all cases we see quite a sharp transition between the percolation type behaviour and the PLER behaviour as $\omega$ varies. 

\begin{figure}
\centering
\includegraphics[width=0.8\linewidth]{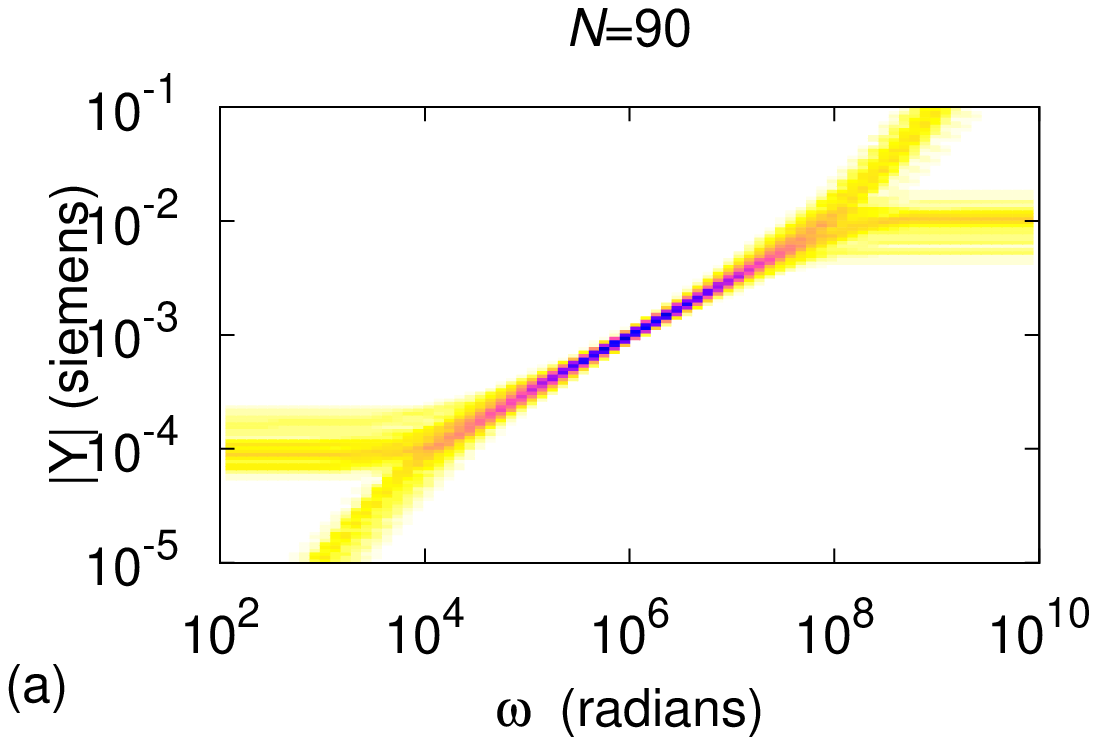}
\includegraphics[width=0.8\linewidth]{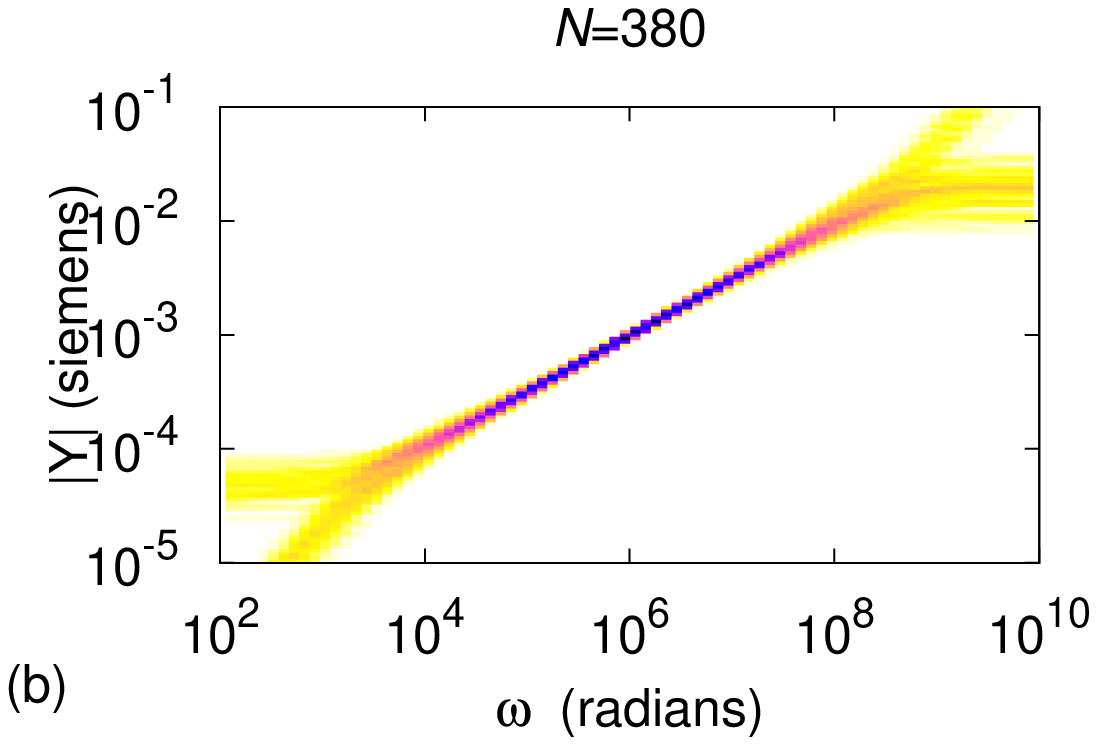}
\includegraphics[width=0.8\linewidth]{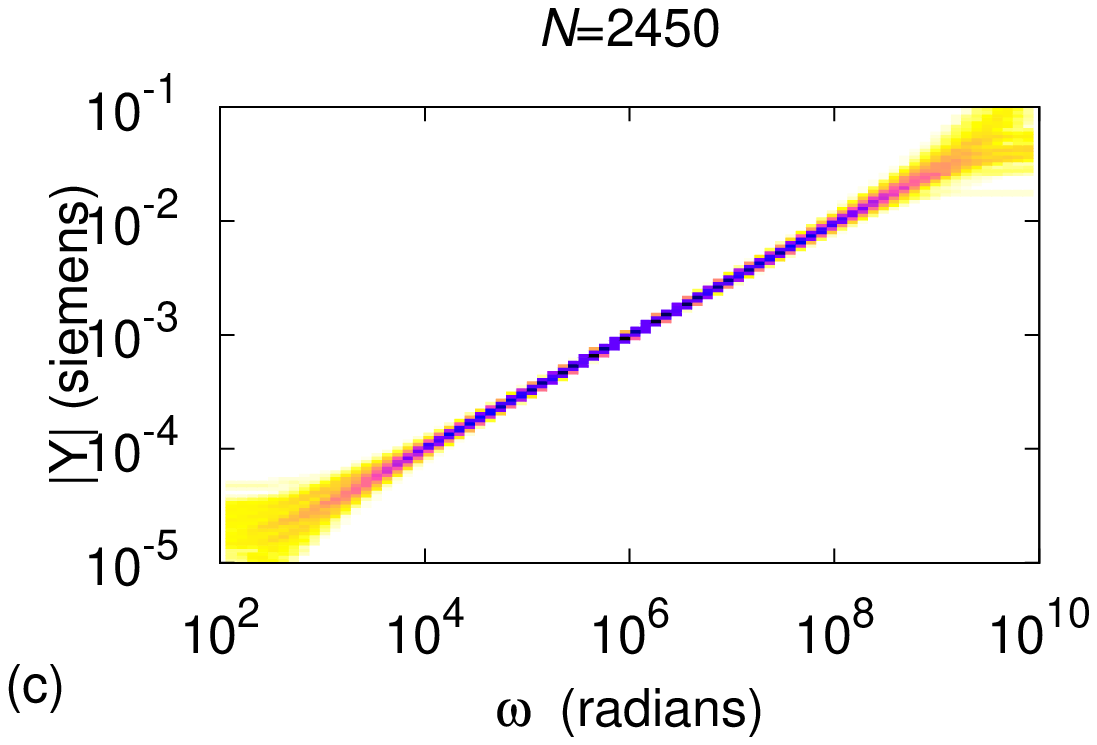}
\includegraphics[width=0.8\linewidth]{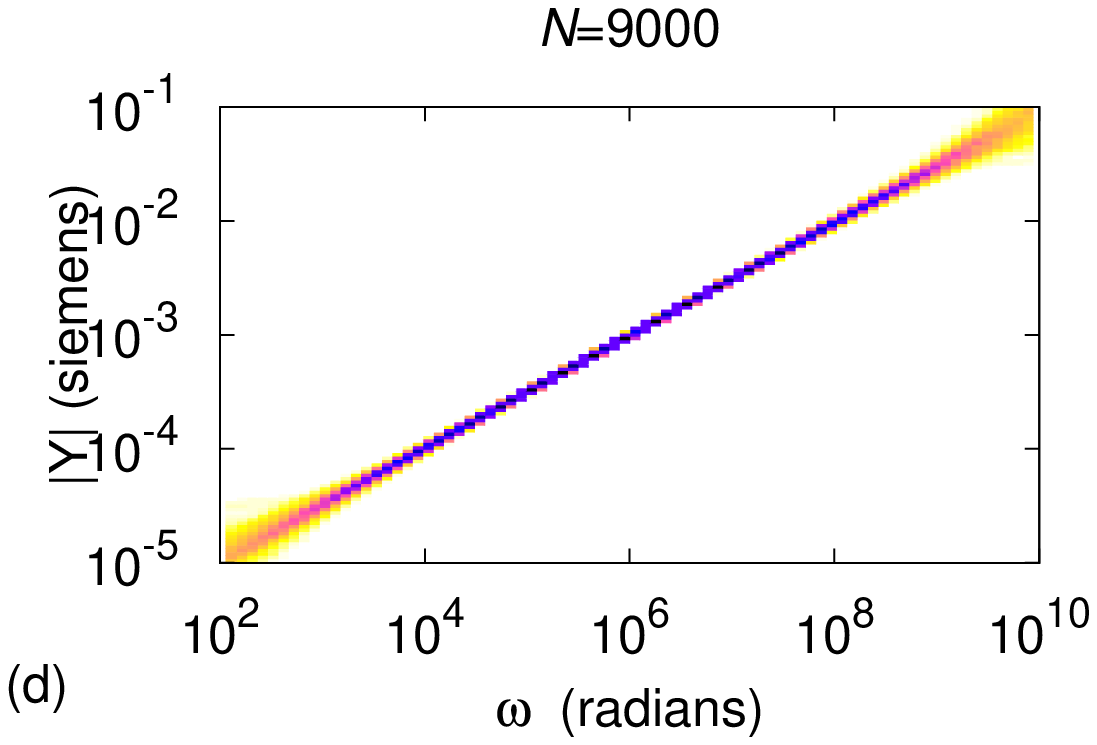}
\caption{(color online) The effect of network size $N$ on the width of the power-law emergent region  for $p=1/2$, in which we see this region increasing without bound as $N$ increases.}
\label{fig:scale50}
\end{figure}

\begin{figure}
\centering
\includegraphics[width=0.8\linewidth]{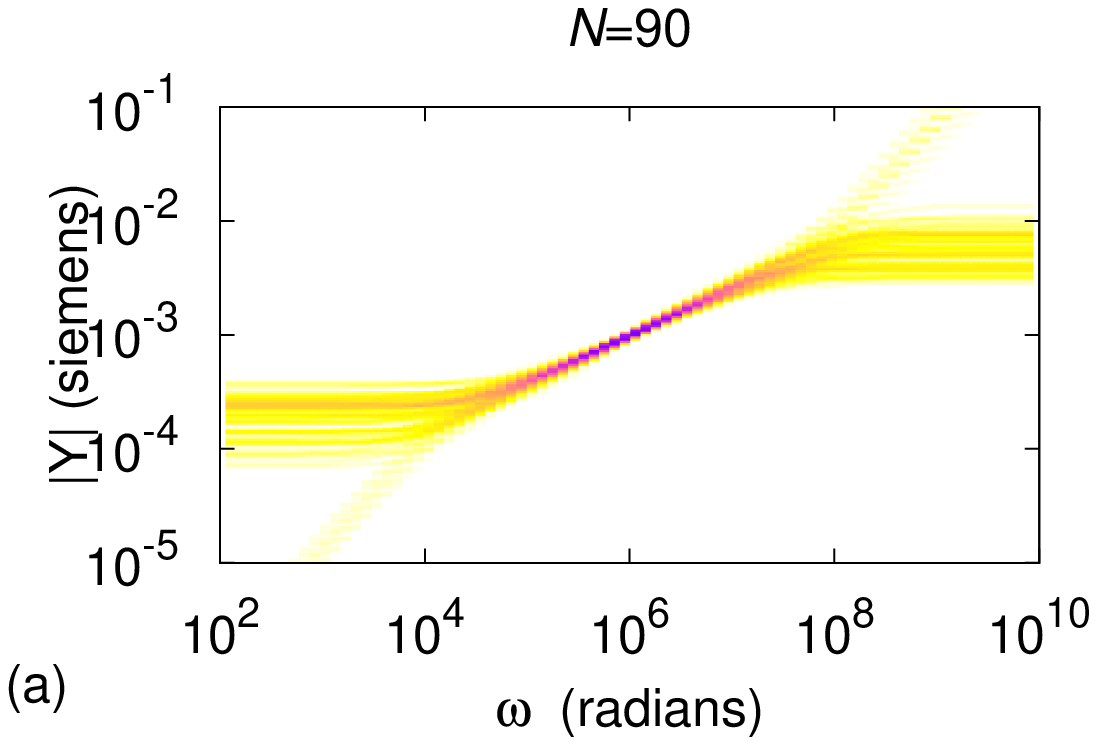}
\includegraphics[width=0.8\linewidth]{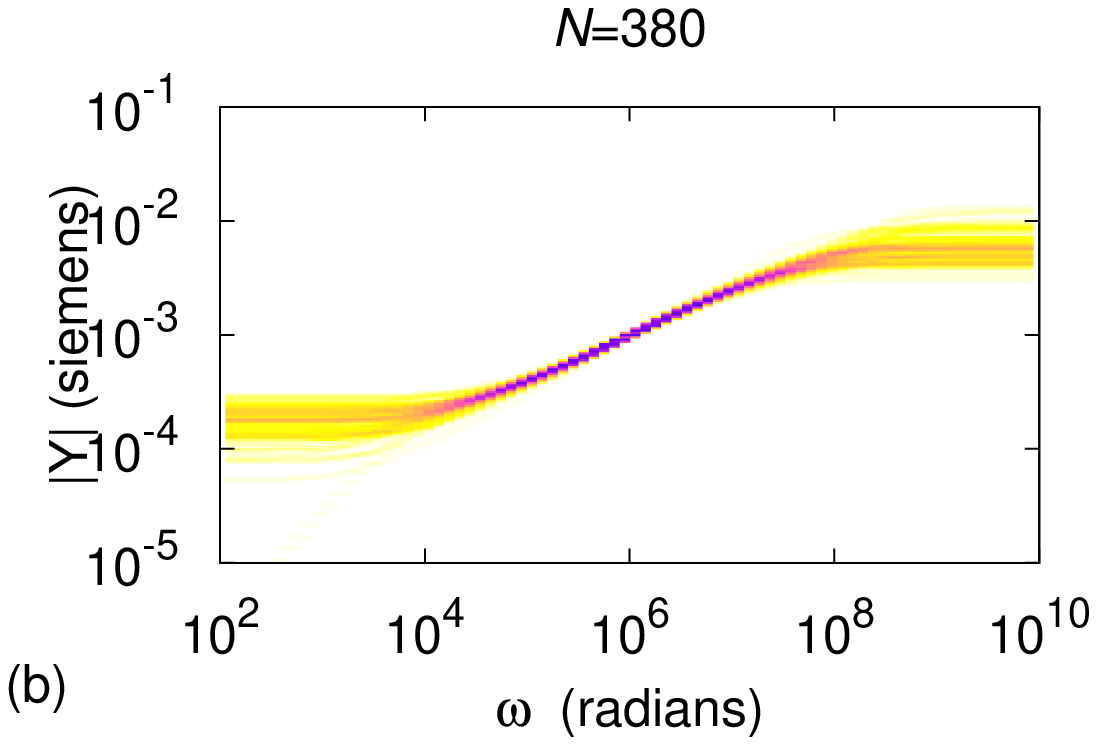}
\includegraphics[width=0.8\linewidth]{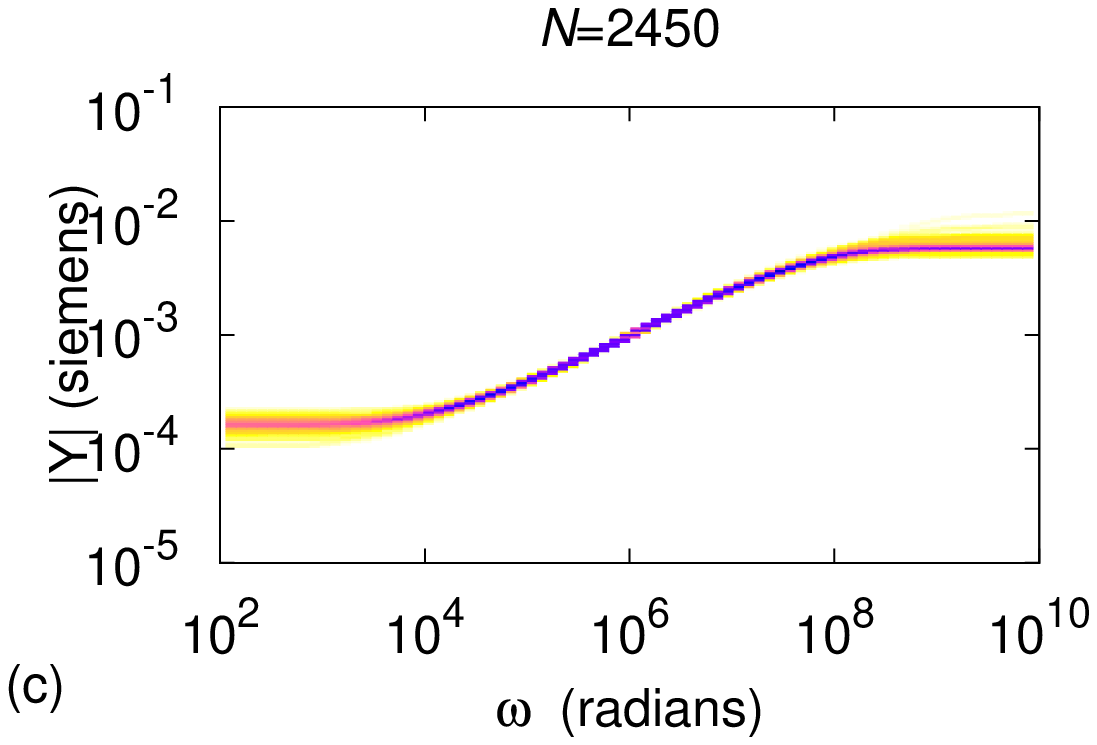}
\includegraphics[width=0.8\linewidth]{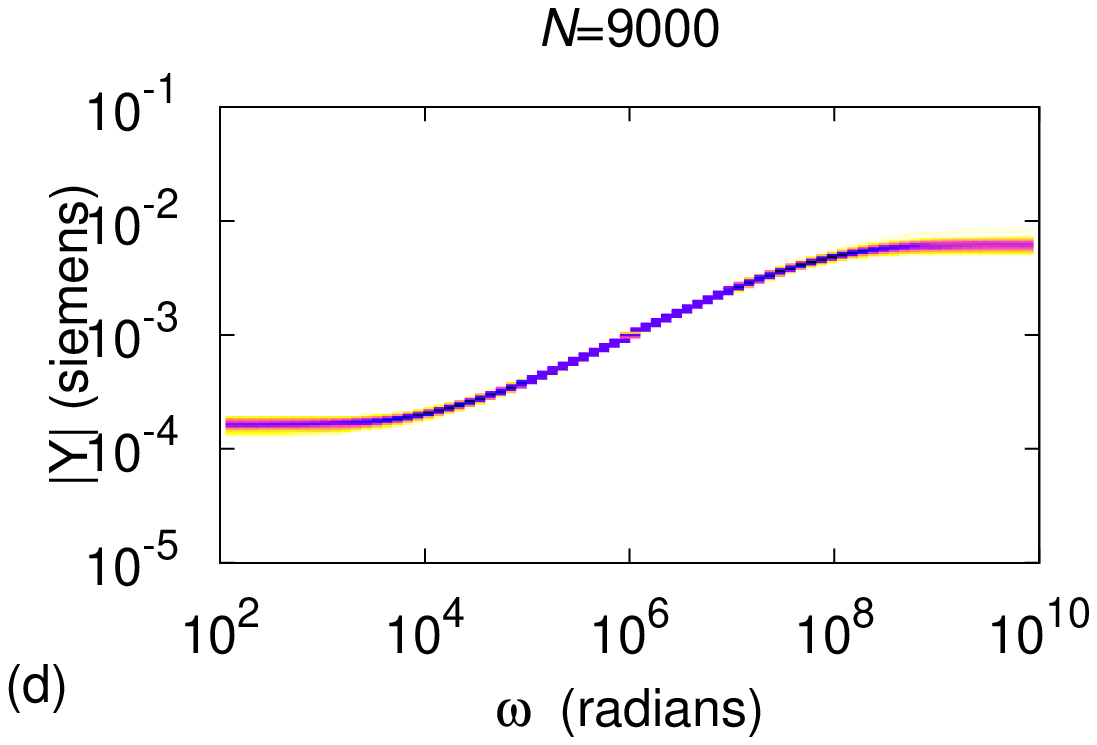}
\caption{(color online) The effect of the network size $N$ on the power-law emergent region for $p=0.4$, in which we see this region becoming asymptotic to a finite set as $N \to \infty$.}  
\label{fig:scale40}
\end{figure}

We have seen above how the response of the network depends strongly upon $p$. It also depends upon the network size $N$, and this effect is critical if $p = 1/2$.
Figure \ref{fig:scale50} shows the response for the critical value of $p=1/2$ for different values of $N$. 
Observe that in this case the width of the power-law emergent region increases apparently without bound, as $N$ increases, as do the magnitude of the responses for small and large frequencies. 
From these graphs, it is apparent that in this critical case the upper limit or the PLER is proportional to $N$ and the lower limit proportional to $1/N$.
We can very roughly motivate the result for $p=1/2$ as follows. Suppose that $\omega$ is small so that the capacitors essentially act as open circuits.  Imagine for a single percolation path through all of these
capacitors  comprising a chain of resistors, then this will have an approximate length of $\sqrt{N}$ resistors and hence a conductance
of $1/(\sqrt{N}R)$. In contrast, if there is a dual  path of capacitors going from top to bottom of the network, interrupting the resistors, then each resitive path has conductance $i\omega C$ and there are
$\sqrt{N}$ of these in parallel, so that the overall conductance is $\sqrt{N} i \omega C$. In Figure \ref{fig:scale40} we plot the response for $p=0.4$ and again increase $N$.
In contrast to the former case, away from $p=1/2$, the size of the power-law emergent
region appears to scale with $N$  for small $N$ before becoming asymptotic to a finite value for larger values of  $N$,  consistent with formul\ae\,(\ref{snow1},\ref{snow2}).


\subsection{The effects of network size and capacitor proportion}

\begin{figure}[ht!]
\centering
\includegraphics[width=0.8\linewidth]{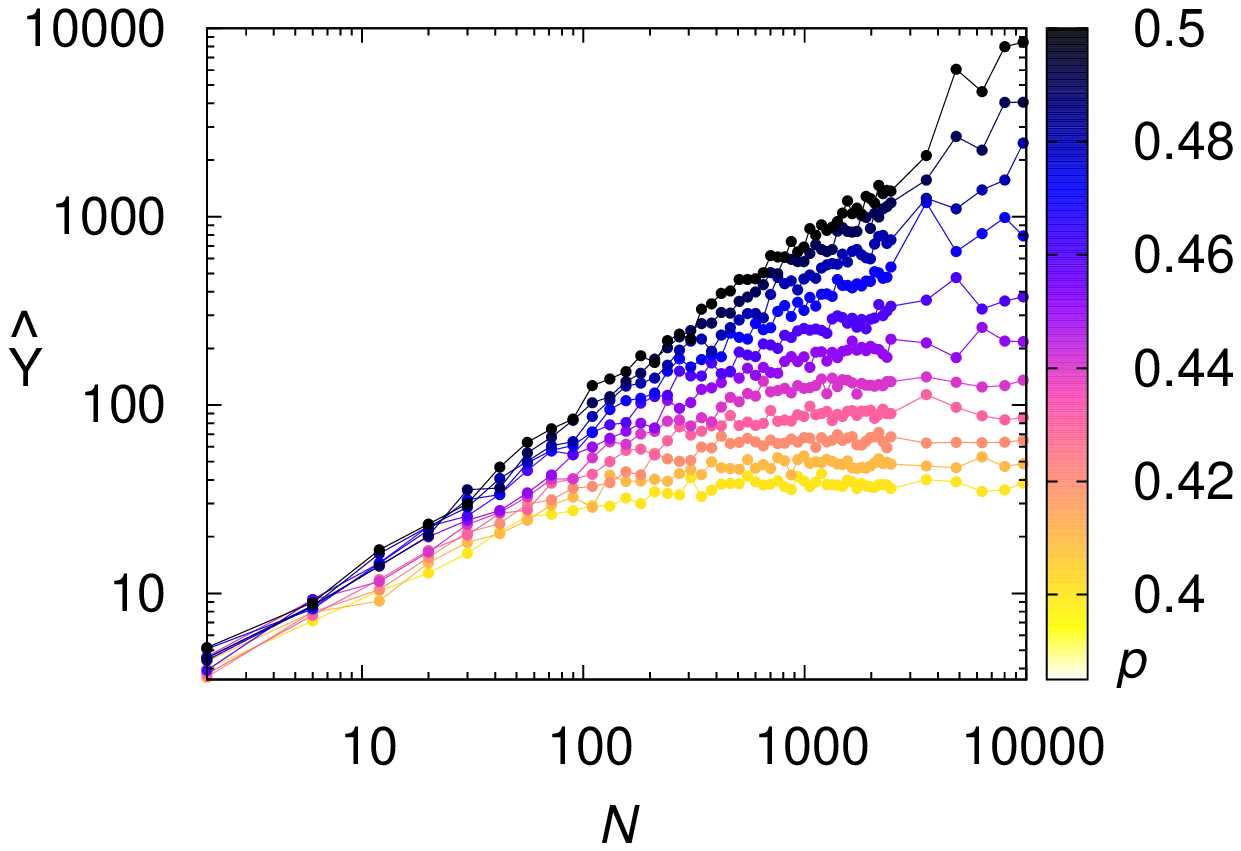}
\includegraphics[width=0.8\linewidth]{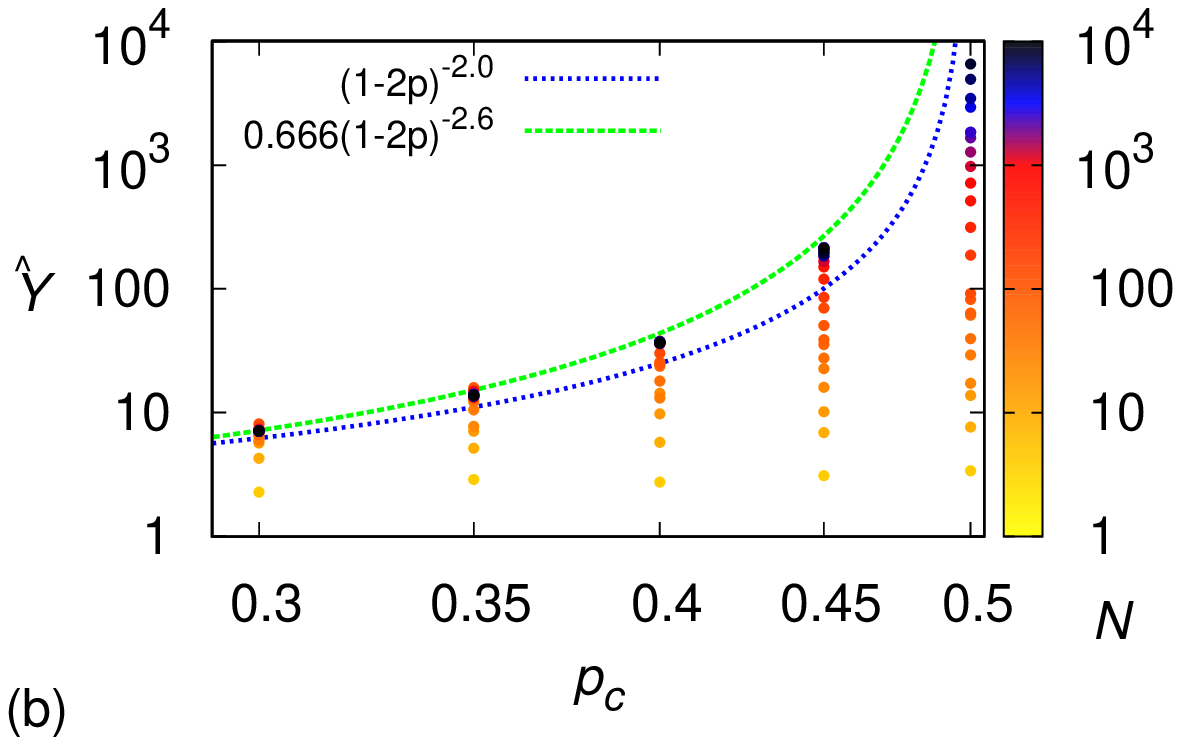}
\caption{(color online) (a) Variation of the dynamic range $\hat{Y} \equiv |Y|_{max}/|Y|_{min}$ as a function of $N$ and $p$,
(b) The value of $\hat{Y}(p)$ as a function of $p \to 1/2$ comparing the estimates
$(1-2p)^{-2}$ and $(1-2p)^{-2.6}$. For each value of $p$ the vertical sequence of dots represents calculations
of $\hat{Y}$ for increasing values of $N$.}
\label{fig:2.6fig}
\end{figure}

To compare these results and to investigate the interplay between network
size and the proportion $p$, we consider for $p \le 1/2$, the {\em dynamic range} of the response for those realisations which have a resistive percolation path  for both low and high frequencies 
(that is with probability one if $ p < 1/2$ and probability $1/4$ if $p = 1/2$). 
We define the dynamic range $\hat{Y}(N,p)$ by
 $$\hat{Y} = \frac{|Y|_{max}}{|Y|_{min}} = \frac{|Y|(\omega \to \infty)}{|Y|(\omega \to 0)}.$$
In Figure \ref{fig:2.6fig} (a)  we plot $\hat{Y}$ as a function of $N$ for a variety of values of $p \le 1/2$.
We see from this figure that if $p = 1/2$ then $\hat{Y}$ is directly proportional to $N$ for all values of $N$. 
In contrast, if $p < 1/2$ then $\hat{Y}$ is directly proportional to $N$ for smaller values of $N$ and then becomes asymptotic to a finite value $\hat{Y}(p)$ as $N \to \infty,$ 
with the asymptotic behaviour occurring when $N > (1/2-p)^{-2}$.
The formul\ae (\ref{rainy1a},\ref{rainy2a}), derived in the final section by combining the EMA and spectral results, imply that $\hat{Y}$ is approximated by $\beta^2$ where $\beta$ satisfies the quadratic equation
\begin{equation}
{
\frac{\beta^2}{N} + (1-2p) \beta - 1 = 0.
}
\label{snow20}
\end{equation}

This gives reasonable qualitative agreement with the calculations presented in Figure \ref{fig:2.6fig} with $\hat{Y} \sim N$ for smaller values
of $N$ and $\hat{Y } \to \hat{Y}(p)$ as $N \to \infty.$ However, we do have to exercise a certain degree of caution in applying this formula. In 
Figure \ref{fig:2.6fig} (b) we present $\hat{Y}(N,p)$ as a function of $p$ as $p \to 1/2$, showing the limiting value $\hat{Y}(p)$ of $\hat{Y}(N,p)$ as $N$ is increased to infinity. We see in this figure that whilst the estimate $\hat{Y}(p) \sim (1-2p)^{-2}$ 
is fairly accurate, a much better estimate in the limit of $p \to 1/2$ is given by
$$\hat{Y}(p) \sim (1-2p)^{-2.6}$$
which is consistent with known empirical results on percolation \cite{jonckheere1998drb}.

\section{Linear circuit analysis}\label{sec:lincirc}

We now describe in detail how the disordered material is modelled by a general electrical network model
with two types of bond of admittance $y_1$ and $y_2$ in respective proportions $1-p$ and $p$. 
These will have admittance ratio $\mu = y_2/y_1.$
For a resistor-resistor (R-R) network with $y_1 = 1/R_1$ and $y_2 = 1/R_2$ we have 
\begin{equation}
\mu = R_1/R_2 \quad \mbox{is real and positive}.
\label{c:oct2}
\end{equation}
For a capacitor-resistor (C-R) network with $y_1 = 1/R$ and $y_2 = i \omega C$  we have
\begin{equation}
\mu = i \omega CR \quad  \mbox{is purely imaginary}.
\label{c:oct3}
\end{equation}
For a capacitor-inductor (C-L) network $\mu$ is a real negative number but we do not consider this case here.
Our interest will be in how the overall admittance of the system varies as the admittance ratio $\mu$ itself varies, and considering how
this can be determined in terms of the poles and zeros of the admittance function $Y(\mu)$.

\subsection{Linear circuit formulation}
\label{sec:poles}

\begin{figure}
\centering
\includegraphics[width=0.8\linewidth]{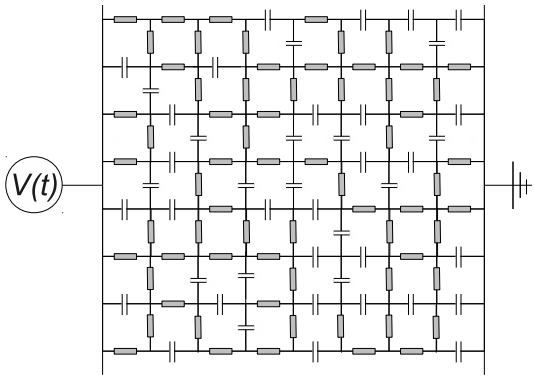}
\caption{(color online) Illustration of an example Resistor Capacitor circuit.}\label{fig:circ}
\end{figure}

Now consider the 2D $N$ node square lattice network shown in Figure \ref{fig:circ}, with all of the nodes on the left-hand-side connected via a bus-bar to a time varying voltage $V(t) = V  e^{i \omega t}$ and on the right-hand-side 
via a bus-bar to earth ($V$). 
 We assign a voltage $v_i$ with $i = 1 \ldots N$ to each (interior) node, and set ${\mathbf v} = (v_1,v_2,v_3 \ldots v_N)^T$.
We also assume that adjacent nodes are connected by a bond of admittance  $y_{i,j} \in \{y_1,y_2\}$. The  current from the node $i$ to an adjoining node at $j$ is then given by $I_{i,j}$ where $I_{i,j} = (v_i - v_j) y_{i,j}.$
From Kirchhoff's current law, at any interior node all currents must sum to zero, so that
\begin{equation}
\sum_{j} y_{i,j} (v_i-v_j) = 0.
\label{c1}
\end{equation}
If $i$ is a node adjacent to the {\em left} boundary then certain of the terms $v_j$ in (\ref{c1}) will take the value of the (known) applied voltage $V(t)$. 
Similarly, if a node is adjacent to the right hand boundary then certain of the terms $v_j$ in (\ref{c1}) will take the value of the ground voltage $0$. 
Combining all of these equations together leads to a system of the form
\begin{equation}
 K{\mathbf v} = V(t) {\mathbf b} = V e^{i \omega t} {\mathbf b},  \label{eqn:mat1a}
\end{equation}
where $K  \equiv K(\omega)$ is the (constant in time) {\em $N \times N$  sparse symmetric Kirchhoff matrix} for the system and the adjacency vector ${\mathbf  b}  \equiv  {\mathbf b}(\omega)$ is the vector of the admittances  of the bonds between the left hand boundary and those  nodes which  connected  to this boundary, with zero entries for all other nodes.
As this is a {\em linear system}, we can take
${\mathbf v} = {\mathbf V} e^{i \omega t}$
so that the (constant in time) vector ${\mathbf V}$ satisfies the linear algebraic equation
\begin{equation}
 K{\mathbf V} = V {\mathbf  b}. \label{eqn:mat1}
\end{equation}

If we consider the total current flow $I$ from the LHS boundary to the RHS boundary then we have
$$I = {\mathbf b}^T (V {\mathbf e} - {\mathbf V})  \equiv Vc - {\mathbf b}^T {\mathbf V},$$
where ${\mathbf e}$ is the vector comprising ones for those nodes adjacent to the left boundary and zeroes otherwise and $c = {\mathbf b}^T{\mathbf e}$. Combining these expressions,  the equations describing the system are then given by
\begin{equation}
{
K {\mathbf V} - {\mathbf b} V  \equiv {\mathbf 0}, \quad c V - {\mathbf b}^T {\mathbf V} = I.
}
\label{dec1}
\end{equation}
The {\em bulk admittance} $Y(\mu)$ of the whole system is then given by $Y = I/V$ so that
\begin{equation}
Y(\mu)= c - {\mathbf b}^T K^{-1} {\mathbf b}.\label{eqn:soly}
\end{equation}

Significantly, the symmetric Kirchhoff matrix  $K$ can be separated into the two sparse symmetric $N \times N$ component matrices $K = K_1+ K_2$  which correspond to the conductance paths along the bonds occupied by each of the two types of components.
Furthermore, using $\mu = y_2/y_1$, we have 
\begin{equation}
K_1 = y_1 L_1 \quad \mbox{and} \quad K_2 = y_2 L_2 = \mu y_1 L_2
\label{c:oct1}
\end{equation}
and hence 
$$K = y_1 L_1 + \mu y_1 L_2,$$
where the terms of the sparse symmetric connectivity  matrices $L_1$ and $L_2$  are constant and take the values $1,0,-1$. 
Note that $K$ is a {\em linear affine function} of $\mu$. 
Furthermore,  
$$\Delta = L_1+L_2$$
is the discrete, positive definite symmetric, negative Laplacian for a 2D lattice. Similarly we can decompose the adjacency vector into two components ${\mathbf b}_1$ and ${\mathbf b}_2$ so that 
$${\mathbf b} = {\mathbf b_1}+{\mathbf b}_2 = y_1 {\mathbf e}_1 + y_2 {\mathbf e}_2 = y_1 {\mathbf e}_1 + \mu y_1 {\mathbf e}_2,$$
where ${\mathbf e}_1$ and ${\mathbf e}_2$ are {\em orthogonal vectors} comprising ones and zeros only corresponding to the two bond types adjacent to the LHS boundary.  Observe again that ${\mathbf b}$ is a linear affine function of $\mu$. 
A similar decomposition can be applied to the scalar $c = y_1 c_1 + \mu y_1 c_2$. 

\subsection{Poles and zeros}

To derive formul\ae\  for the expected admittances in terms of the admittance ratio $\mu$ we now examine the structure of the admittance function $Y(\mu)$.
As the matrix $K$, the adjacency vector ${\mathbf b}$  and the scalar $c$ are all affine functions of the parameter $\mu$ it follows immediately from (\ref{eqn:soly}) and Cramer's rule applied to (\ref{dec1}) that the admittance of the network $Y(\mu)$ is rational function of the parameter $\mu$, taking the form of the ratio of two complex polynomials $P(\mu)$ and $Q(\mu)$ of respective degrees $r \le N$ and $s \le N$, so that
\begin{equation}
 Y(\mu)=\frac{Q(\mu)}{P(\mu)}=\frac{q_0+q_1\mu+q_2\mu^2+ \ldots q_r \mu^r}{p_0+p_1\mu+p_2\mu^2+\ldots p_s \mu^s}.
 \label{c:oct4}
 \end{equation}
We require that $p_0 \ne 0$ so that  the response is physically realisable, with $Y(\mu)$ bounded as $\omega$ and hence $\mu \to 0$.
Several properties of the network can be immediately deduced from this formula. For convenience, we look at a C-R network, although similar results arise for R-R networks. 
First consider the case of $\omega$  {\em small}, so that $\mu = i \omega CR$ is also small. 
From the discussions in Section \ref{sec:models}, we predict that either there is  (a) a resistive percolation path in which case $Y(\mu) \sim \mu^0$ as $\mu \to 0$ or (b) such a path does not exist, so that the conduction is capacitative with $Y(\mu)  \sim \mu$ as $\mu \to 0$.
The case (a) arises when $p_0 \ne 0$ and the case (b) when $q_0 = 0$. Observe that this implies that the absence of a resistive percolation path as $\mu \to 0$ is equivalent to the polynomial $Q(\mu)$ {\em having a zero when $\mu = 0$}. 
Next consider the case of $\omega$ and hence $\mu$ large. 
In this case
$$Y(\mu) \sim \frac{q_r}{p_s} \mu^{r-s} \quad \mbox{as} \quad \mu \to \infty.$$
This time we may have (c) no capacitive path at high frequency with response $Y(\mu) \sim \mu^0$ as $\mu \to \infty$, or the existence of a capacitative path with $Y(\mu) \sim \mu$. 
In case (c) we have $s = r$ and $p_r \ne 0$ and in case (d) we have $s = r-1$ so that we can think of taking $p_r = 0$. 
Accordingly, we identify four types of network defined in terms of the percolation paths for low and high frequencies, which correspond to the cases (a), (b), (c), (d) so that
\begin{center}
\begin{tabular}{|c|c|}
\hline
(a)	& $p_0 \ne 0$\\ \hline
(b)	& $p_0 = 0$\\ \hline
(c)	& $p_r  \ne 0$\\ \hline
(d)	& $p_r = 0$\\ \hline
\end{tabular}
\end{center}
Both the polynomials $P(\mu)$ and $Q(\mu)$ can be factorised by determining their respective roots $\mu_{p,k}$, $k=1 \ldots s$ and $\mu_{z,k}$, $ k=1 \ldots r$ which are the {\em poles} and \emph{zeroes} of $Y(\mu)$. 
We will collectively call these poles and zeroes the \emph{resonances} of the network. Our analysis of the network will rely on determining certain statistical and other properties of these resonances. 
Note that in Case (b) we have $\mu_{z,1} = 0$.  Accordingly the network admittance can be expressed as
\begin{equation}
Y(\mu,N) = D(N) \frac{\prod\limits_{k=1}^r (\mu - \mu_{z,k})}{\prod\limits_{k=1}^s (\mu - \mu_{p,k})}.\label{eqn:prod}
\end{equation}
Here $D(N)$ is a function which does not depend on $\mu$ but does depend on the characteristics of the network.  

\subsection{Location of the resonances}

We now proceed to prove some rigorous results concerning the location of the poles and zeroes.
We firstly note that the number $r$ of poles/zeroes can be substantially less than $N$. This is due to the formation of {\em clusters} of components in the lattice which are isolated from the boundaries \cite{clerc1990ecb}.
Such component clusters lead to resonances at infinity or zero, depending on which component the clusters are made of.
Comparing (\ref{eqn:soly}) and (\ref{eqn:prod}), it can be seen immediately that the poles are precisely the roots of the determinant of the Kirchhoff
matrix $K$.
This matrix has the form $K = K_1 + K_2 = y_1 (L_1 + \mu L_2)$ with $L_1$, $L_2$ constant and symmetric (though not necessarily positive-definite) and $L_1+L_2 = \Delta$. 
The poles are then given by  $-1$ times the eigenvalues of the matrix pencil $(L_1,L_2)$, so that
the values $\mu_{p,k}$, and the corresponding vectors ${\mathbf v}_{p,k}$, satisfy the linear equation
\begin{equation}
\label{eq:gen_evp}
(L_1 + \mu_{p,k} L_2){\mathbf v}_{p,k} = 0 \quad \text{with} \quad {\mathbf v}_{p,k} \neq 0.
\end{equation}
As $L_1 + L_2 = \Delta$ this then implies that
$$(L_1(1- \mu_{p,k}) + \mu_{p,k} \Delta){\mathbf v}_{p,k} = 0$$ so that $$(L_1 + \mu_{p,k} \Delta/(1 - \mu_{p,k})){\mathbf v}_{p,k} = 0.$$

It follows immediately from the  symmetry of $L_1$ and the fact that $\Delta$ is a symmetric positive definite operator, that
$\mu_{p,k}/(1 - \mu_{p,k})$ is real. The negativity of $\mu_{p,k}$ follows from the fact that the network has a bounded response.

In contrast, using (\ref{eqn:soly}) and (\ref{dec1}), the zeros are those values of $\mu = \mu_{z,k}$, with  corresponding vectors ${\mathbf v}_{z,k}$ which satisfy the simultaneous equations
$$(L_1 + \mu_{z,k} L_2) {\mathbf v}_{z,k} = {\mathbf b}, \quad {\mathbf b}^T {\mathbf v}_{z,k} = c,$$
so that the current $I$ is zero.
The condition for the zero can thus be put into an extended  matrix equation  of the form
\[
\left[\begin{matrix}
 K & -{\mathbf b}  \\
 -{\mathbf b}^T &  c
\end{matrix}\right]
\left[\begin{matrix}
{\mathbf v_{z,k}} \\
1
\end{matrix}\right]
= 0.
\]
As $K$, ${\mathbf b}$ and $c$ are all affine functions of $\mu$, this leads to the corresponding eigenvalue problem for the zeros $\mu_{z,k}$ given by
\begin{equation}
\label{eq:gen_evp_ext}
\left[\begin{matrix}
y_1(L_1 + \mu_{z,k} L_2)     &         y_1(  -{\mathbf e}_1-\mu_{z,k} {\mathbf e}_2)\\
y_1(-{\mathbf e}_1^T-\mu_{z,k} {\mathbf e}_2^T )       &y_1(   c_1+\mu_{z,k} c_2)
\end{matrix}\right]{\mathbf v}_{z,k} = 0,
\end{equation}
with ${\mathbf v}_{z,k}\neq 0$, so that the zeros are the eigenvalues $\mu_{z,k}$ of this extended matrix pencil.
Both problems (\ref{eq:gen_evp}) and (\ref{eq:gen_evp_ext}) can be transformed into standard eigenvalue problems.
From the previous reasoning, problem (\ref{eq:gen_evp}) is equivalent to the problem
\begin{equation}
L_1 {\mathbf v}_{p,k} = \frac{\mu_{p,k}}{\mu_{p,k}-1} \Delta{\mathbf v}_{p,k}
\label{eq:L1}
\end{equation}
and using the fact that $\Delta$ is a symmetric positive definite matrix, the Cholesky decomposition \cite{GolubvanLoan96} $\Delta= LL^T$ exists, where $L$ is a lower triangular matrix.
Thus we can rewrite (\ref{eq:L1}) as a standard eigenvalue problem
\begin{equation}
\label{eq:std_evp}
L^{-1} L_1L^{-T} {\mathbf w_{p,k}} = \zeta_{p,k} {\mathbf w}_{p,k},\quad\text{with}\quad {\mathbf w_{p,k} } \neq 0,
\end{equation}
where ${\mathbf w}_{p,k} = L^T {\mathbf v}_{p,k}$ and $\zeta_{p,k} = \dfrac{\mu_{p,k}}{\mu_{p,k}-1}$. Note that as $\mu_{p,k} < 0$ it follows that
$0 < \zeta_{p,k} < 1.$
Similarly,  (\ref{eq:gen_evp_ext}) can be written as
\[
\left[\begin{matrix}
 L_1 & -{\mathbf e}_1\\
 -{\mathbf e}_1^T&  c_1
\end{matrix}\right] {\mathbf v_{z,k}} = \frac{\mu_{z,k}}{\mu_{z,k}-1} \left[\begin{matrix}
(L_1 +  L_2)  &  -{\mathbf e}_1-{\mathbf e}_2\\
 -{\mathbf e}_1^T- {\mathbf e}_2^T & c_1+ c_2
\end{matrix}\right]{\mathbf v_{z,k}}.
\]
Using the Cholesky decomposition of $\Delta $ again we can define an extended generalised Cholesky decomposition of the extended matrix system by
\begin{align*}
&\quad\, \left[\begin{matrix}
L_1 +  L_2    &  -{\mathbf e}_1-{\mathbf e}_2\\
 -{\mathbf e}_1^T- {\mathbf e}_2^T     &  c_1+ c_2
\end{matrix}\right]\\
&= \left[\begin{matrix}
L & 0\\
 -({\mathbf e}_1 + {\mathbf e}_2)^T L^{-T}&  \alpha
\end{matrix}\right]\left[\begin{matrix}
L^T& -L^{-1}({\mathbf e}_1 + {\mathbf e}_2)\\
0&  \alpha
\end{matrix}\right]\\
 &= \hat{L}\hat{L}^T,
\end{align*}
where $\alpha^2 =c_1+c_2 - ({\mathbf e}_1+{\mathbf e}_2)^T(LL^T)^{-1}({\mathbf e}_1+{\mathbf e}_2)$ and $\hat{L}$ is a lower triangular matrix. Using this extended Cholesky decomposition we rewrite (\ref{eq:gen_evp_ext}) as
\begin{equation}
\label{eq:std_evp_ext}
\left[\begin{matrix}
L^{-1}  L_1L^{-T}& \hat{{\mathbf b}}\\
\hat{{\mathbf b}}^T&  \hat{c}
\end{matrix}\right]  {\mathbf w}_{z,k}
= \zeta_{z,k} {\mathbf w}_{z,k},\quad\text{with}\quad {\mathbf w}_{z,k} \neq 0,
\end{equation}
${\mathbf w}_{z,k} = \hat{L}^T {\mathbf v}_{z,k}$, $\zeta_{z,k} =
\displaystyle\frac{\mu_{z,k}}{\mu_{z,k}-1}$, $\hat{\mathbf{b}} =(L^{-1}L_1 L^{-T}({\mathbf e}_1+{\mathbf e}_2)-L^{-1}{\mathbf e}_1))/\alpha$ and $\hat{c} = (({\mathbf e}_1+{\mathbf e}_2)^T L^{-1}L_1 L^{-T}({\mathbf e}_1+{\mathbf e}_2)-{\mathbf e}_1^T L^{-T}({\mathbf e}_1+{\mathbf e}_2)-({\mathbf e}_1+{\mathbf e}_2)^T L^{-1}{\mathbf e}_1 +c_1))/\alpha^2$.

Now, by the {\em Cauchy interlacing theorem} \cite[Theorem 10.1.1]{Parlett98} (see also \cite{Wilkinson65}), the eigenvalues of (\ref{eq:std_evp}) interlace those of (\ref{eq:std_evp_ext}), that is
\[
0 \le \zeta_{z,1} \le \zeta_{p,1} \le \zeta_{z,2}  \le \zeta_{p,2} \le \ldots \le \zeta_{p,s} \le \zeta_{z, (s+1)} \le 1.
\]
Furthermore the eigenvalues $\zeta_{z,k}, i=1,\ldots,s+1$ corresponding to the zeros are given by the zeros of the function
\[
f(\lambda) = \lambda-\hat{c}+\sum_{k=1}^r\frac{|\hat{{\mathbf b}}_k|^2}{\zeta_{p,k}-\lambda}.
\]

Equivalently, the poles and zeros of $Y(\mu)$ are
all negative real numbers, and interlace so that
\begin{align}
0 \ge \mu_{z,1} \geq  \mu_{p,1} &\geq \mu_{z,2} \geq \mu_{z,2} \ldots\nonumber\\
 &\geq \mu_{p,s} (\geq \mu_{z,s+1}).
\label{interlace:1}
\end{align}

This result immediately leads to two different interpretations in the case of an R-R and a C-R network.
In the case of an R-R network with conductance ratio $\mu > 0 $ the poles and zeros occur along the
{\em negative real axis} so that we can take $\mu_{p,k} = -M_{p,k} < 0$ etc. Thus, as $\mu$ varies through positive real values
\begin{equation}
Y(\mu) = D(N) \frac{\prod_{k=1}^r (\mu + M_{z,k})}{\prod_{k=1}^s (\mu +M_{p,k})},
\label{cbfreq2}
\end{equation}
with the values $M_k^z \ge 0$ and $M_k^p > 0 $.
For the C-R network,  $\mu = i \omega CR$, and we can consider $Y$ to be a function of $\omega$. The poles $\omega_{p,k}$ of $Y(\omega)$ then satisfy
$i CR \omega_{p,k} = -M_{p,k}$ so that they lie along the positive imaginary axis, ditto the zeros. Thus,  as $\omega$ varies through real values then
\begin{equation}
Y(\omega) = D(N) \frac{\prod_{k=1}^r (\omega - i W_{z,k})}{\prod_{k=1}^s (\omega - i W_{p,k})},
\label{cbfreq1}
\end{equation}
with the values $W_k^z \ge 0$ and $W_k^p > 0 $. We note that neither of the expressions (\ref{cbfreq1}, \ref{cbfreq2}) become unbounded as $\omega$ varies through real values or as $\mu$ varies through positive real values.
This is in contrast to the case of an $C-L$ network in which the resonances can be real and positive  can lead to unbounded responses as $\omega$ varies.
In contrast, we see in the $C-R$ and $R-R$ networks, an averaging effect in the product terms in these expressions, which leads to the  emergent behaviours observed in practice.

\section{The distribution of the resonances}\label{sec:distributions}

We now look at the distribution of the poles and zeros, and draw certain conclusions about their
statistical regularity which will allow us to then compute the asymptotic form
of the system response. The statistics of the resonances are most
regular in the critical case of $p=1/2$, allowing us to 
make very precise estimates of the overall system behaviour in this case, precisely complementing the averaging 
methods which work best when $p \ne 1/2$. 
To perform these calculations, we note that if we consider the elements of the network to be assigned randomly, with the components taking each of the two possible values with probabilities $p$ and $(1-p)$,
then we can consider the resonances to be random variables. The poles and associated eigenvectors are given by the solutions of the matrix pencil equation (see (\ref{eq:gen_evp}))
\begin{equation}
{
L_1 {\mathbf v}_{p,k} = -\mu_{p,k} L_2 {\mathbf v}_{p,k}.
}
\label{bath101}
\end{equation}
Each realisation of the network, with bonds chosen from a Binomial $[p, (1-p)]$ distribution
will give a different set of values for $\mu_{p,k} \equiv -i M_{p,k} \equiv i CR W_{p,k} $ and we can then consider the statistics of this set.
We ask the following questions:  (1) What is the statistical distribution of  $\mu_{p,k}$ if $N$ is large? (2) What is the statistical distribution of the location of a zero between its two
adjacent poles? (3) How do $\mu_{p,1}$ and $\mu_{p,N}.$ 
In each case we will find good numerical evidence for strong statistical regularity of the poles (especially in the case of $p = 1/2$), leading to answers to each of the above questions. 

\subsection{Preliminary observations on the pole locations}

\begin{figure}
\centering
\includegraphics[width=0.8\linewidth]{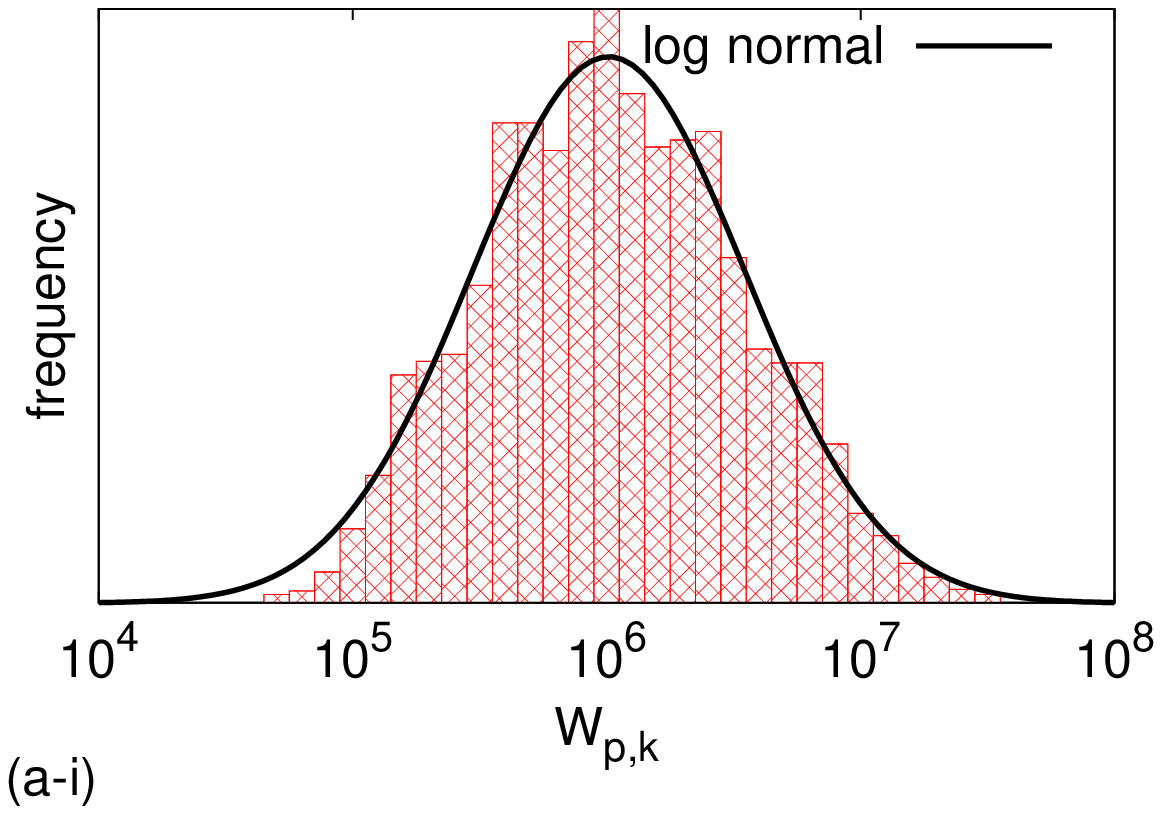}
\includegraphics[width=0.8\linewidth]{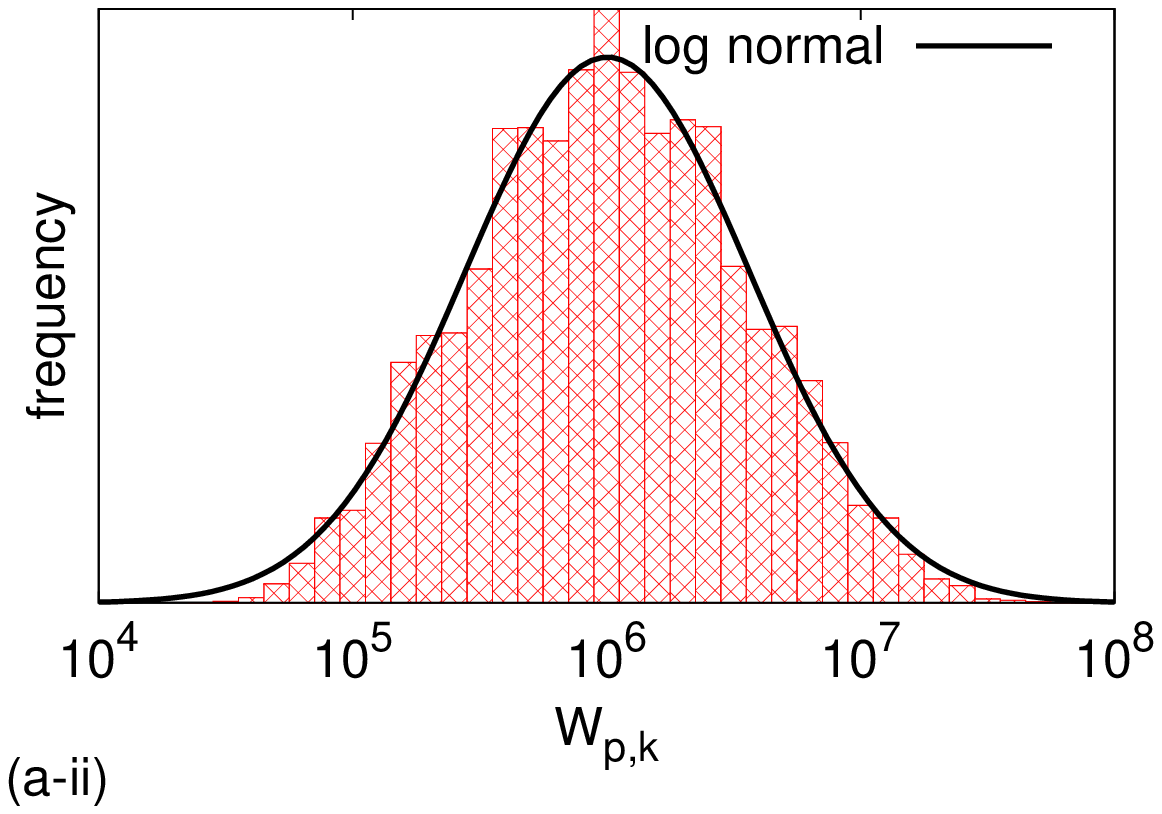}

\includegraphics[width=0.8\linewidth]{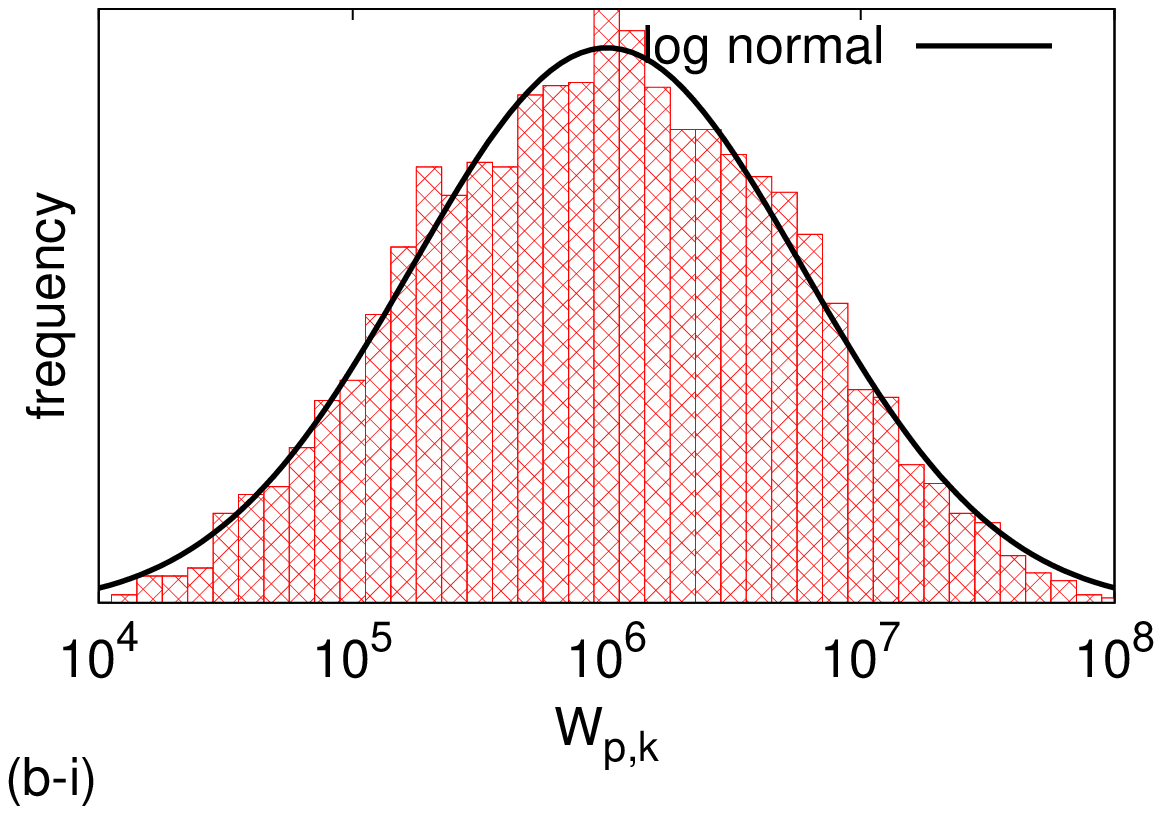}
\includegraphics[width=0.8\linewidth]{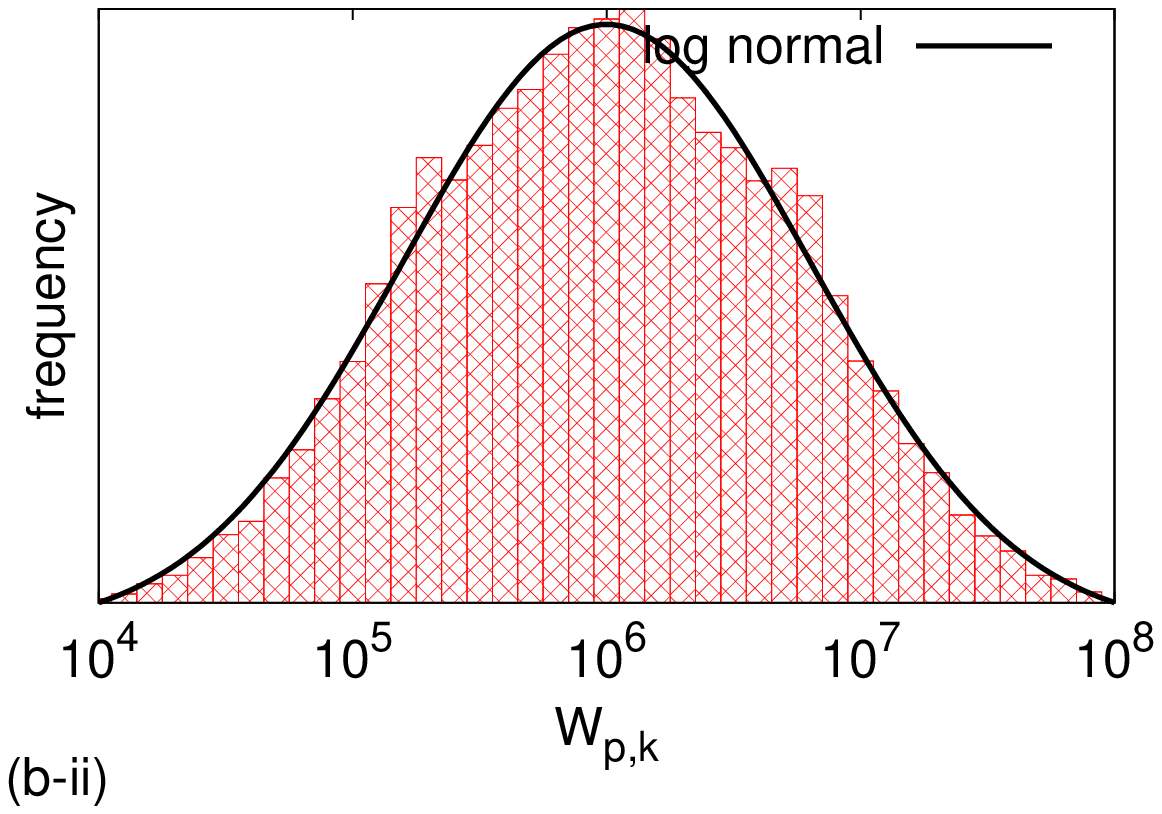}
\caption{(color online) Distribution of $W_{p,k}$ values for (a) $p=0.25$ and (b) $p=0.5$ over a range of system size: (i) $N=90$, (ii) $N=380$.  The curves are fitted to Normal distributions (on a log-scale). The variance appears to depend on $p$ but be largely independent of the value of $N$.}\label{fig:fit}
\end{figure}

\begin{figure}[ht!]
\centering
\includegraphics[width=0.8\linewidth]{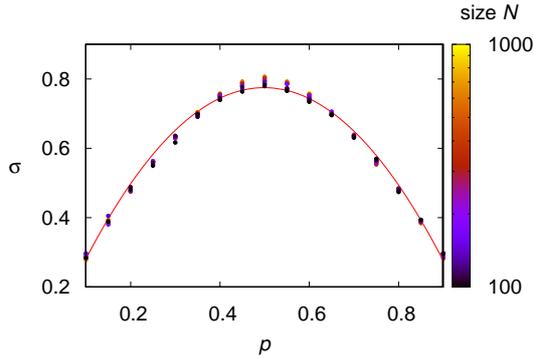}
\caption{(color online) The standard deviation $\sigma$ as a function of $p$ showing $\sigma \approx \alpha p(1-p)$.}\label{fig:powersig}
\end{figure}

\begin{figure}[hb!]
\centering
\includegraphics[width=0.8\linewidth]{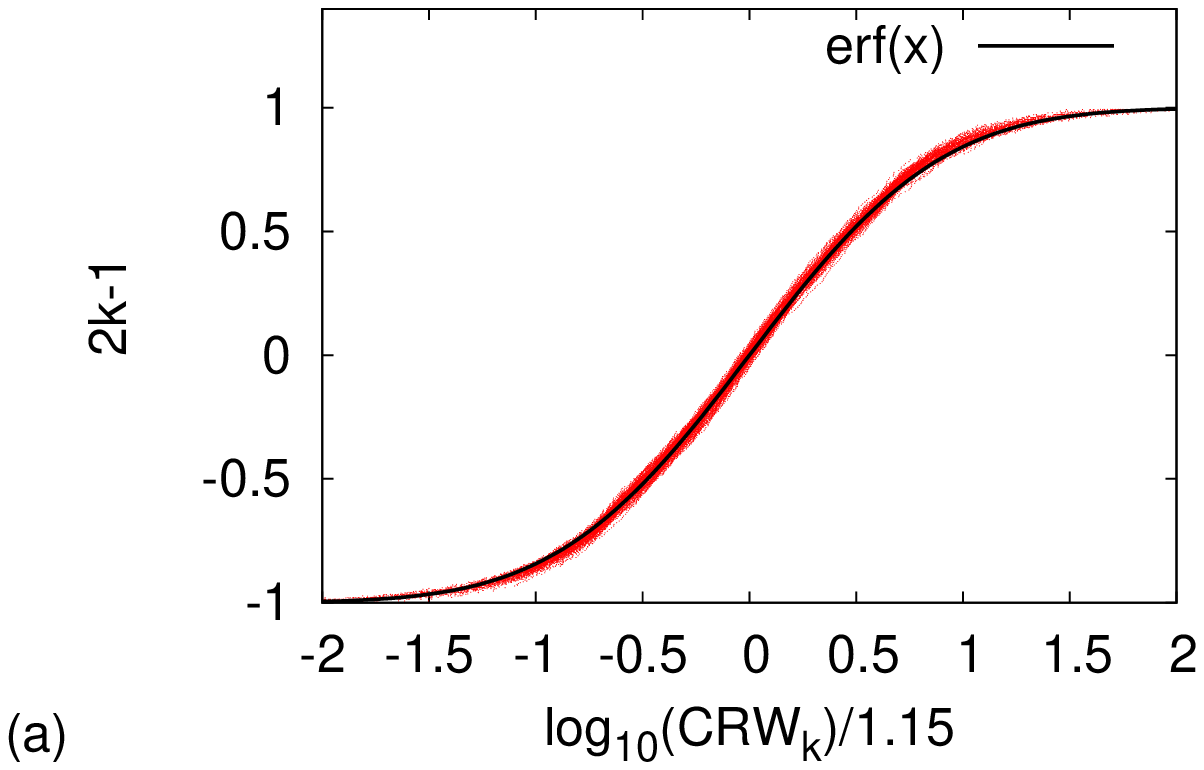}
\includegraphics[width=0.8\linewidth]{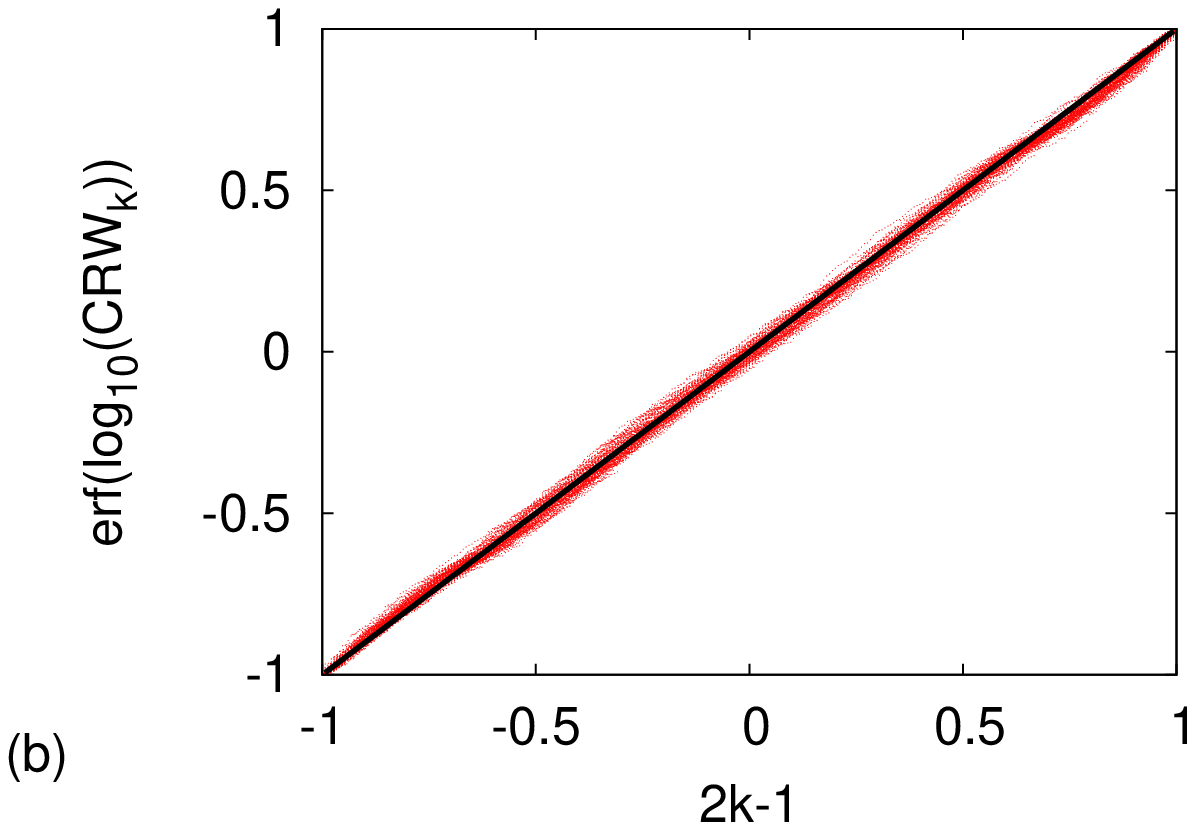}
\caption{(color online) The location of the logarithm of the poles as a function of $k$, for a single realisation of the network,  and a comparison with the inverse error function.}\label{polelocs}
\end{figure}

To motivate our answers we start by considering  the special case of $p = 1/2$. 
In this case the two matrices $L_1$  and $L_2$ representing the connectivity of the two components have a statistical duality.
Statistically, any realisation which leads to a particular matrix $L_1$ is equally likely to lead to the same matrix $L_2$. 
Because of this, if $\mu$ is an observed eigenvalue of the pair $(L_1, L_2)$ then it is equally likely for there to be an observed eigenvalue of the pair $(L_2, L_1)$
with the same eigenvector and with eigenvalue being precisely $1/\mu$. 
Thus in any set of realisations of the system we expect to see the eigenvalues $\mu$ and $1/\mu$ occurring with equal likelihood. 
It follows from this simple observation that the variable $\log(\mu)$ should be expected to be a random variable with a symmetric probability distribution and with mean zero. 
It is therefore natural to expect that for a large number of realisations,
the variables $\log(M_{k,p})$ should follow a normal distribution with mean zero (so that $M_{p,k}$ has a {\em lognormal} distribution centred on $M = 1$). 
Similarly, if $M_{p,1}$ is the smallest value of $M_{p,k}$ and $M_{p,N}$ the largest value then $M_{p,1} = 1/M_{p,N}$. 
In fact we will find that in this case of $p=1/2$ we have $M_{p,1} \sim 1/N$ and $M_{p,N} \sim N$. 
It follows similarly that $\log(W_{p,k})$ is expected to have a mean value of $-\log(CR)$.
Following this initial discussion, we now consider some  numerical computations of the distribution of the poles in a C-R network for which $CR = 10^{-6}$.
As a first computation we consider many random realisations of networks  generated with a large enough size (typically $N = 380$) to ensure good statistics per network.
We define $S$ as the number of horizontal components in one row of the network; giving $S^2$ horizontal and $(S-1)^2$ vertical components.
The number of internal nodes (i.e.\ not including the boundary nodes), which is equal to the dimension of the matrix $K$, is therefore $N=S (S-1)$; giving the maximum possible number of eigenvalues $\mu_i$.
The results of the computations are presented in Fig.~\ref{fig:fit} in which we give a histogram of the distribution of the poles $W_{p,k}$ (on a log-scale in the frequency domain) over 100 different realisations of each network.
These figures clearly indicate that the location of the poles does indeed possess a strong statistical regularity, conforming approximately to a {\em log-normal distribution} with mean $\log(1/CR)$  in all cases.
Evidence for this is given by comparing the resulting curve with the standard Normal distribution with an appropriately chosen value for the variance.
The fitted curves in Figure~\ref{fig:fit} show that the results are close to log-normal for any choice of $p$ (provided that $N$ is chosen sufficiently large).
When the results of the realisations considered above are fitted to a log-Normal distribution with probability density function $P(W) = a \exp \left( -(W-E\{W\})^2/2\sigma^2 \right) $ we find the remarkable result that the standard deviation $\sigma$ appears to be largely {\em independent of the value of $N$} and to display a simple functional relation on $p$. 
In Figure~\ref{fig:powersig} we show the standard deviation as a function of $\sigma$ for a range of values of $N$ and see
a good fit to the curve $\sigma = \alpha p(1-p)$
over many values of $N$.
As a second calculation we take a {\em single realisation} of a network with $N \approx 380$ nodes and $p = 1/2$ and determine the location of the imaginary part of poles $W_{p,k}$.
A plot of the logarithm of the poles, ordered in increasing size, as a function of $k$ is given in Figure~\ref{polelocs}. 
Two features of this figure are immediately  obvious. 
Firstly, the terms $W_{p,k}$ appear to be point values of a regular function $f(k)$. 
Secondly,  $\log(CR W_{p,k})$ shows a strong degree of symmetry about zero, so that if $1 \le  k \le N$ then   $\log(CR W_{p,k}) = 0$ if $k = N/2$.
Motivated by the discussion above,
we compare the form of this graph with that of the {\em error function}, that is we compare $\mbox{erf}(\log(CRW_{p,k}))$ with $2k/N - 1$. 
The correspondence is very good, strongly indicating that $\log(f)$ takes the form of the inverse error function with an appropriate constant of proportionality.

\subsection{Pole-zero spacing}

As a next calculation we consider the statistical distribution of the location of the interlacing zeros with respect to the poles. 
In particular we consider the variable $\eta_k$, which depends on the proportion $p$ given by
\begin{equation}
\eta_k (p)\equiv   \frac{\log M_{p,k+1} - \log M_{z,k}}{\log M_{p,k+1} - \log M_{p,k}} \equiv  \frac{\log W_{p,k+1} - \log W_{z,k}}{\log W_{p,k+1} - \log W_{p,k}}.
\label{chrisdelta}
\end{equation} 
We now establish three symmetry results for the mean value $\bar{\eta}_k(p)$ of $\eta_k$, taken over many realisations.

%%% NDS: begin insert
\underline{First symmetry}: Assume the zeros are $W_{z,k},k\in[0,N]$ and the poles are $W_{p,k},k\in[1,N]$. Define,
\begin{align}
\hat{\eta}_k(p) &\equiv \frac{\log \hat{W}_{p,k+1} - \log \hat{W}_{z,k}}{\log \hat{W}_{p,k+1} - \log \hat{W}_{p,k}},
\end{align}
where,
\begin{align}
\log \hat{W}_{z,k} &= - \log W_{z,N-k},\\
\log \hat{W}_{p,k} &= - \log W_{p,N-k+1}.
\end{align}
So,
\begin{align}
\hat{\eta}_{N-k}(p) &\equiv \frac{\log W_{z,k} - \log W_{p,k}}{\log W_{p,k+1} - \log W_{p,k}}.
\end{align}
Hence,
\begin{align}
\hat{\eta}_{N-k}(p) + \eta_k(p) &= 1.
\end{align}
It follows from simple symmetry considerations that
$L_1(p)$ has the same form on average as $L_2(1-p)$ and vice versa. Hence,
if $\mu_{p,k}$ is observed for one realisation with $y_2$ in proportion
$p$, then $1/\mu_{p,k}$ will be observed when the proportion of $y_2$
is $1-p$. A similar result holds for the zeros.
Noting that,
\begin{align}
\bar{\hat \eta}_{N-k}(p) &= \bar{\eta}_{N-k}(1-p),
\end{align}
then,
\begin{equation}
{
\bar{\eta}_k(p) + \bar{\eta}_{N-k}(1-p) = 1.
}
\label{bath102}
\end{equation}

\begin{figure}
\centering
\includegraphics[width=0.8\linewidth]{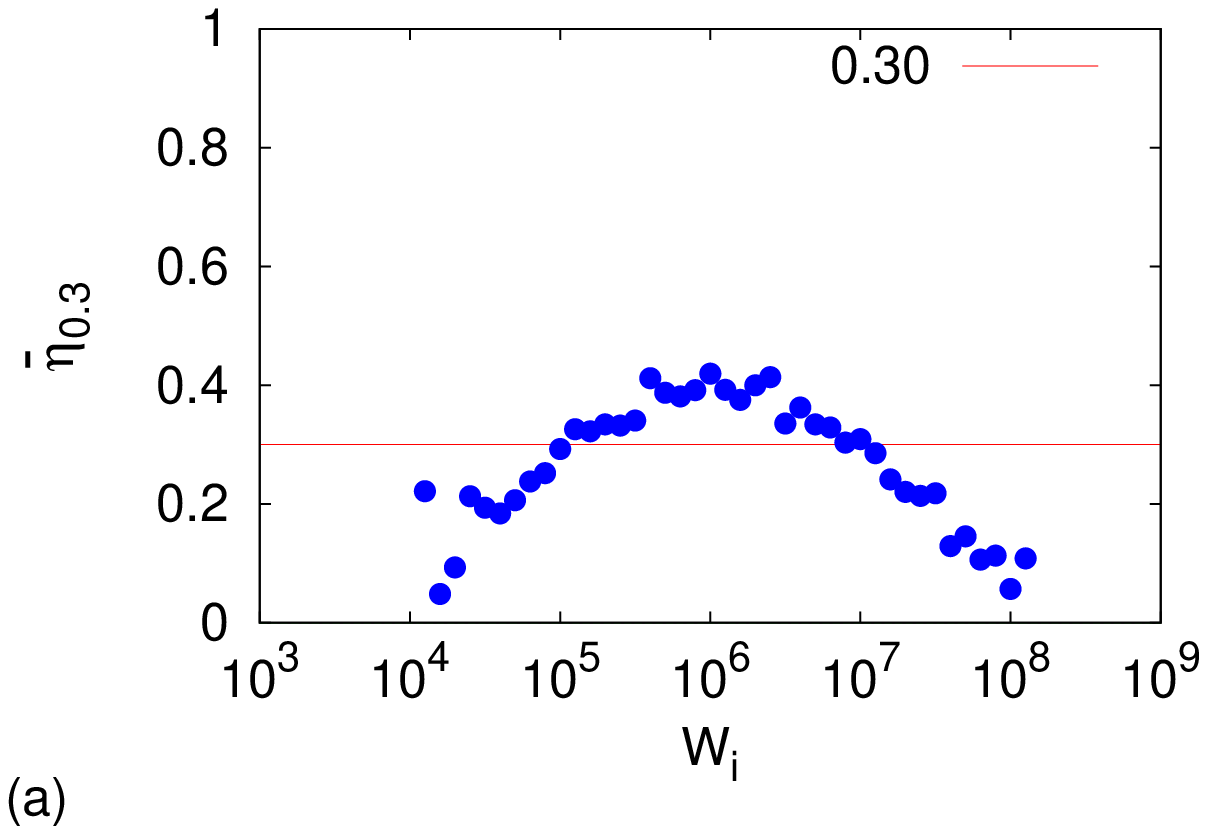}
\includegraphics[width=0.8\linewidth]{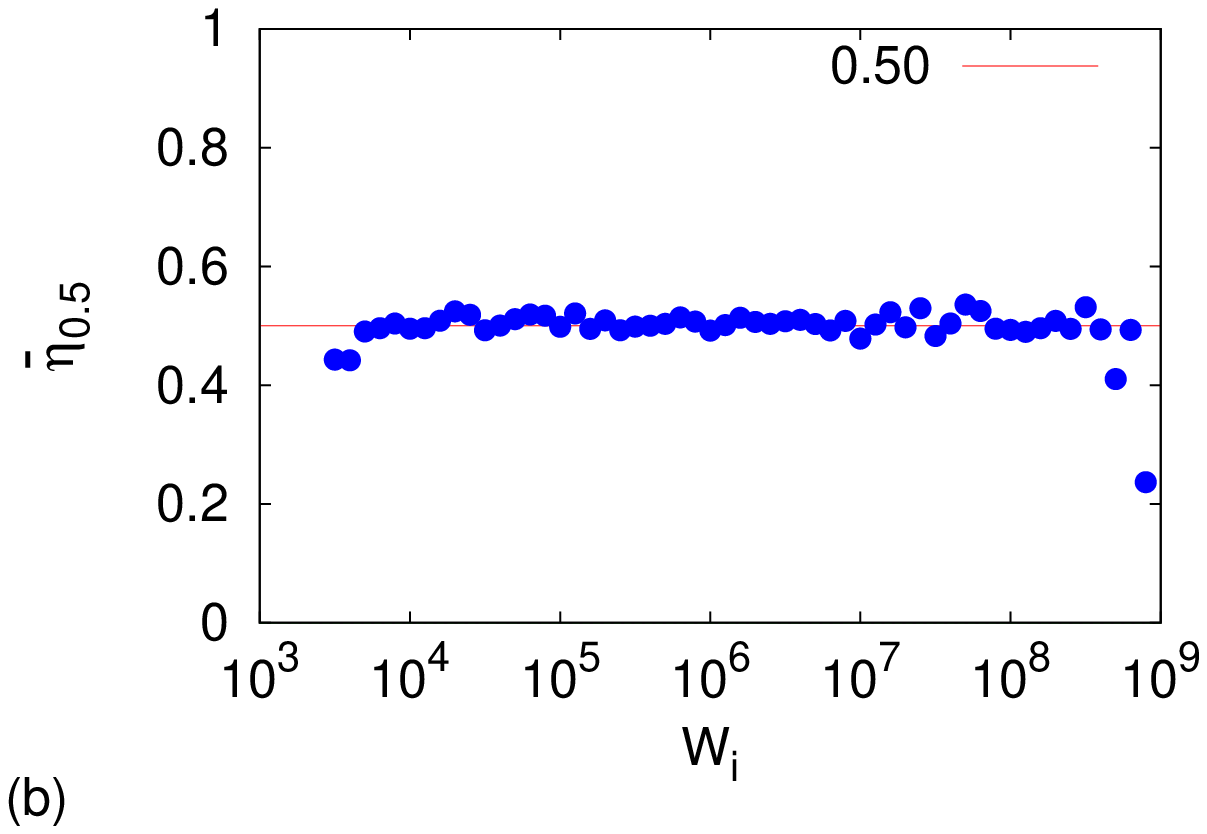}
\includegraphics[width=0.8\linewidth]{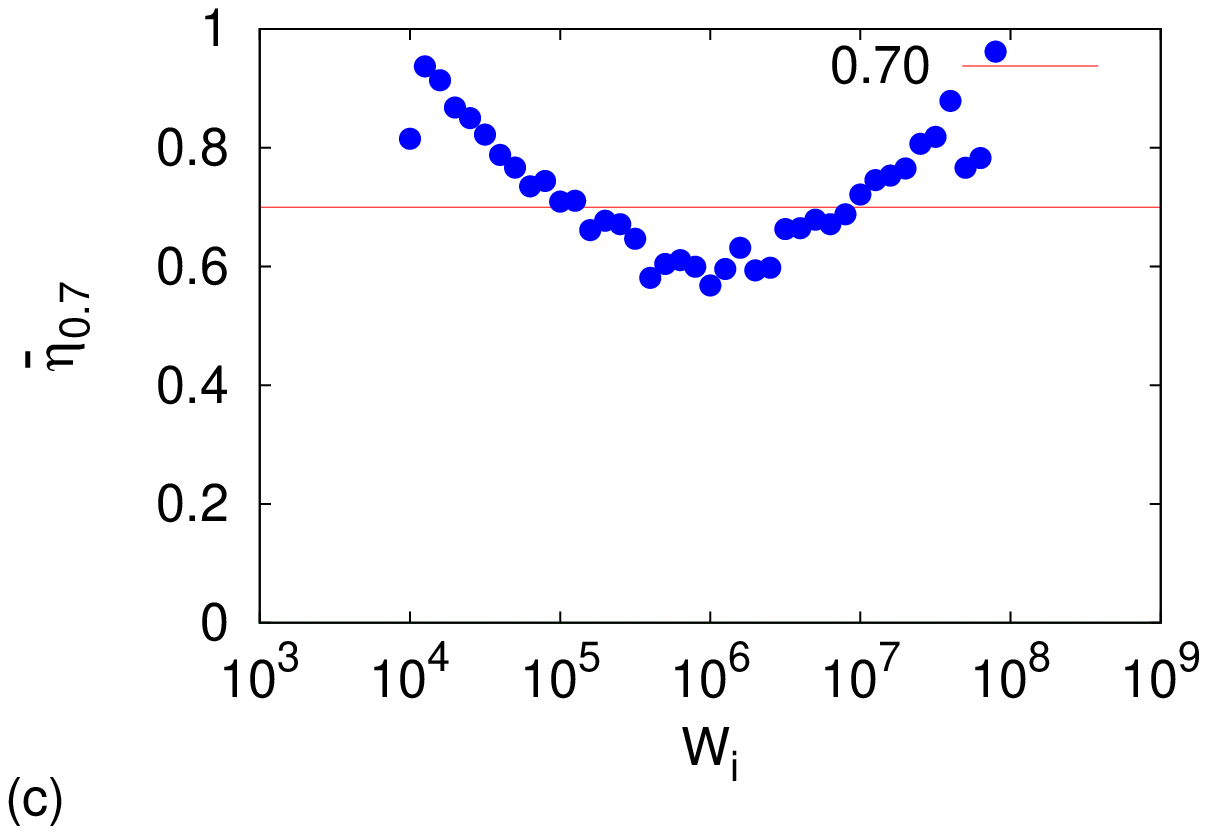}
\includegraphics[width=0.8\linewidth]{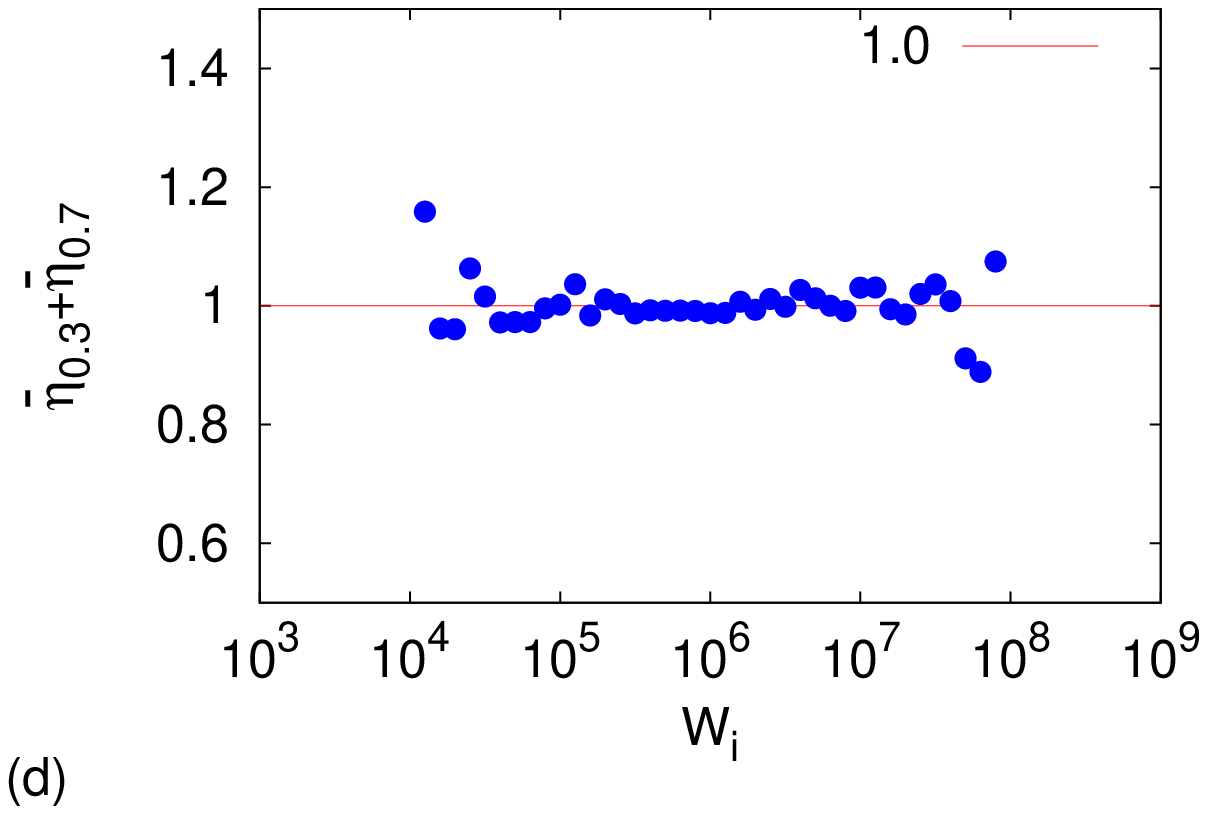}
\caption{(color online) Figures showing how the mean value $\bar{\eta}_k$ of $\eta_k$ taken over many realisations of the network, varies with the mean value of $W_{p,k}$.  The four examples show results for (a) $p=0.3$, (b) $p=0.5$, (c) $p=0.7$ and (d)  $\bar{\eta}(0.3)+ \bar{\eta}(0.7)$.}\label{fig:rangedif}
\end{figure}

\underline{Second symmetry}: As a second observation we invoke  duality results due to Keller \cite{milton1980bcd} (see also \cite{jonckheere1998drb}), in which the admittance of a network is compared with that of the dual network, in which every bond of the original network is replaced with an orthogonal bond for the dual. 
Significantly, square binary networks are self-dual. 
A consequence of the duality results is that if
\begin{equation}
Y(y_1,y_2) \; Y(y_2,y_1) = y_1 \; y_2.
\label{dual}
\end{equation}
In the case where $p=0.5$ it is often argued that as it claimed that as $Y(y_1,y
_2)= Y(y_2,y_1)$ then $Y = \sqrt{y_1,y_2}$.
As we have seen in Section \ref{sec:models}, this is only correct in the PLER (where the response is unique). 
It this is not quite correct in the percolation region where $Y$ can take one of two forms, but it does correctly predict the duality between these two forms.
It follows from (\ref{dual}) that 
$$D(N) \frac{\prod_{k=1}^r (\mu + M_{z,k})}{\prod_{k=1}^s (\mu +M_{p,k})} = \frac{y_1 y_2}{D(N)} \; \frac{\prod_{k=1}^s (1/\mu + M_{p,k})}{\prod_{k=1}^r (1/\mu+M_{z,k})}.$$
This can only be true for all $\mu$ if we have the symmetry result (taking the ordering of the poles and zeros into account) given by
$$M_{p,k} = 1/M_{z,N-k}.$$
It immediately follows from (\ref{chrisdelta}) that asymptotically we have the second symmetry
\begin{equation}
{
\bar{\eta}_k(p)  =  \bar{\eta}_{N-k}(p).
}
\label{csymm}
\end{equation}
\underline{Third symmetry}: Combining (\ref{bath102}) and (\ref{csymm}), we derive the third symmetry
%%% NDS: end of insert
\begin{equation}
{
\bar{\eta}_k(p) + \bar{\eta}_k(1-p) = 1.
}
\label{csymm2}
\end{equation}
In particular, this gives
\begin{equation}
{
\bar{\eta}_k(1/2) = 1/2.
}
\label{csymm1}
\end{equation}
The distribution of $\bar{\eta}_k(p)$ (over 100 {\em realisations} of a C-R network) plotted as a function the location of $\log(W_{p,k})$ for $p = 0.3, 0.5, 0.7$,
is shown in Figure \ref{fig:rangedif} together with a graph of $\bar{\eta}_k(0.3) + \bar{\eta}_k(0.7)$.
The  figures (a),(b) and (c) show clearly the reflectional symmetry about the mid-point implied by (\ref{csymm}).
The figure in part (b) (with $p=1/2$) is particularly remarkable, clearly indicating, as predicted by (\ref{csymm1}) that   $\bar{\eta}_k(1/2) $ is equal to $1/2$ almost independently of the value of $\log(W_{p,k})$. 
There is some deviation from this value at the high and low ends of the range due to slower convergence to the mean.
As well as this there is some evidence for a small asymmetry in the results, but the constancy of the mean near to 1/2 is very convincing.  
The figure in part (d) for $p = 0.3$ and $p = 0.7$ clearly illustrates the symmetry relation (\ref{csymm2}).
We note, however, that for $p \ne 1/2$  the value of $\bar{\eta}_k$ of $\eta_k$ varies with $\log(W_{p,k})$ in a symmetric distribution (as predicted by (\ref{csymm})) which depends approximately quadratically on the value of $\log(W_{p,k})$. 
If $p > 1/2$, $\bar{\eta}_k$ takes a value a little less than $p$ in the centre of the range when $W_{p,k} = W_{mid} = 1/CR$,  and a bit greater than $p$ towards the ends of the range. 
The distribution is reversed when $p < 1/2$, as can be seen by comparing Figures \ref{fig:rangedif} (a) and (c), and this is a consequence of (\ref{csymm2}).

\subsection{Limits of the resonance distributions}

\begin{figure}[ht!]
\centering
\includegraphics[width=0.8\linewidth]{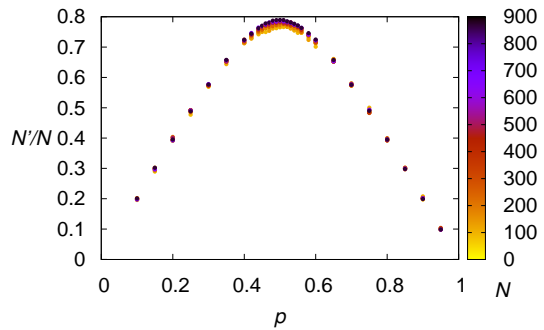}
\caption{(color online) The value of $N'/N$ for varying values of $p$ and $N$.}\label{Npvals}
\end{figure}

\begin{figure}[hb!]
\centering
\includegraphics[width=0.8\linewidth]{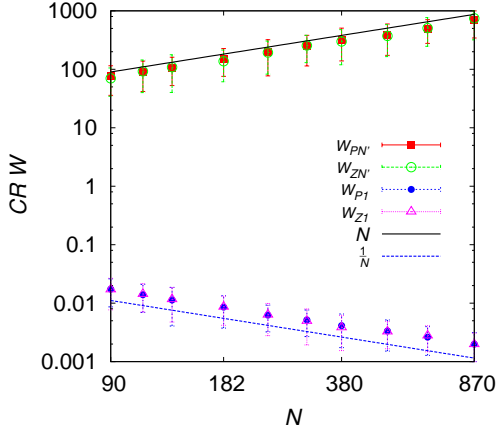}
\caption{(color online) Largest $W_{z,N'}, W_{p,N'}$ and smallest $W_{z,1},W_{p,1}$ zeros and poles for a C-R network with $p=0.5$ showing linear dependence on $N$ and $1/N$.}\label{fig:minmax}
\end{figure}

As a final calculation, we consider the number $N'$  of the {\em finite non-zero resonances} in this case of a C-R network, and the location of the first non-zero pole and zero $W_{z,1},W_{p,1}$ and the last finite pole and zero $W_{p,N'},W_{z,N'}$. 
As discussed, in the case of $p=1/2$ we expect a symmetrical relation so that $CR W_{p,1}$ and $ CR W_{p,N'}$ might be expected to take reciprocal values.
We consider two calculations, firstly determining  $N'/N$ for a range of values of $N$ and of $p$ and secondly calculating the functional dependence of $W_{p,1}$ and $W_{p,N'}$ upon $N$ and $p$.The value of $N'$ can be considered statistically and represents probability of a node contributing to the current paths. 
If we take $z = N'/N$ as a function of $p$ for a range of values of $N$ the resulting distribution is plotted in Figure~\ref{Npvals}. We see that the shape of this curve is parabolic in $p$ with a maximum value for $z \approx 0.8$ given when $p = 1/2$.
Indeed, statistical arguments presented in \cite{jonckheere1998drb} indicate that the maximum value at $p = 1/2$ is given by
$N' = 3 \left( 2 - \sqrt{3} \right) = 0.804 \ldots  .$
We next consider the values of $W_{p,1}$ and of $W_{p,N'}$ which will mark the transition between emergent type behaviour and percolation type behaviour. 
A log-log plot of the values of $W_{z,1}$, $W_{p,1}$ and of $W_{z,N'}$, $W_{p,N'}$ as functions of $N$ for the case of $p = 1/2$  is given in Figure~\ref{fig:minmax}. 
There is very clear evidence from these plots that each of $W_{z,1}$, $W_{p,1}$ and $W_{z,N'}$, $W_{p,N'}$ both have a {\em strong linear dependence} upon $N$ and $1/N$  for {\em all values of $N$}. 
Indeed we conclude from this figure that the following reciprocal relations hold
$$CR \; W_{z,1}, \; CR \; W_{p,1}  \sim N^{-1}$$ and $$CR \; W_{z,N'}, \; CR \; W_{p,N'}  \sim N,$$
with an identical scaling for $M_{z,1},M_{p,1},M_{z,N'},M_{p,N'}$.

\subsection{Summary}

The main conclusions of this section are that there is a strong statistical regularity in the location of the poles and the zeros of the admittance function. 
In particular we may make the following conclusions based on the calculations reported in this section.

\begin{enumerate}
\item $M_{p,k} \sim f(k)$ for an appropriate continuous function $f(k)$ where $f$ depends upon $p$ strongly and upon $N$ very weakly.
\item $\bar{\eta}_k(1/2) \approx 1/2$ for all values of $k$.
\item If $p = 1/2$ and if $M_{z,1} \ne 0$, then 
$M_{p,1},M_{z,1} \sim N^{-1}, \quad M_{p,N'},M_{z,N'} \sim N.$
\end{enumerate}

\section{Asymptotic analysis using spectral results}\label{sec:asymp}

We now use the results in the summary of the previous section to derive the  form of the conductance in the two cases of a C-R network and an R-R network.
The results are sharp for the case of $p = 1/2$ and $N < \infty$ allowing us to deduce the asymptotic responses in this case.
The formul\ae\  that we derive will take one of four forms, depending upon the nature of the percolation paths.

\subsection{Derivation of the response for general $\mu$}

We consider the formul\ae\  for the value of the admittance  of the binary network
\begin{equation}
 Y(\mu)= D(N) \frac{ \prod \limits_{k=1}^{r} (\mu - \mu_{z,k})}{\prod \limits_{k=1}^{s} (\mu - \mu_{p,k})},
\label{eqn:trigged}
\end{equation}
where the results of section (\ref{sec:lincirc}) imply that $\mu_{z,k} = -M_{z,k}$ , and 
$0 \le M_{z,1} < M_{p,1} < M_{z,2} < M_{p,2} <    \ldots < M_{p,s}  ( < M_{z (s+1)}).$
Crucially, as $\mu$ is either positive or purely imaginary, $Y$ is a bounded function for all $\mu$.
Here we assume that we have $s = N'$ poles, but consider situations with different percolation responses for $|\mu|$ large or small, depending upon whether the first zero  $M_{z,1} = 0$ and on the existence or not of a final zero $M_{z,(N'+1)}.$ 
These four cases lead to four functional forms for the conductance, all of which are realisable in the case of $p = 1/2$. 
In this section we derive each of these four forms from some simple asymptotic arguments. 
At this stage the constant $D(N)$ is undetermined, but we will be able to deduce its value from our subsequent analysis. 
Although simple, these arguments lead to remarkably accurate formul\ae\  when $p = 1/2$, when compared with the numerical calculations, that predict not only the PLER but also the limits of this region.
To obtain an asymptotic formula from (\ref{eqn:trigged}) we assume that $s = N'$ is large, and that there is a high density of poles and zeros.
From the results in the previous section we know that, asymptotically, the poles at $-M_{p,k}$ follow a regular distribution and that the the zeroes have a regular spacing between the poles, especially in the case of $p = 1/2$. 
The conclusions of the previous section on the distribution of the poles and the zeros leads to the following formul\ae:
\begin{align}
M_{p,k} &\sim f(k),\nonumber\\
 \frac{M_{p,k+1}-M_{z,k+1}}{M_{p,k+1}-M_{p,k}} &=  \delta_k, \nonumber\\
 M_{p,k+1}-M_{p,k} &\sim f'(k),\nonumber\\
 M_{z,k+1} &\sim f(k) + (1-\delta_k) f'(k)\label{bigasy}.
\end{align}
Here, as we have seen, the function $\log(f(k))$ is given by the inverse of the error function, but its precise form does not matter too much for the next calculation.
To do this we firstly consider the contributions to the product in (\ref{eqn:trigged}) which arise from the terms from the first pole to the final zero, so that we consider the following product:
\begin{equation}
P \equiv D(N) \prod\limits_{k=1}^{N'} \frac{\mu + M_{z,k}}{\mu + M_{p,k}}.\label{eqn:trigged1}
\end{equation}
Note that this product has implicitly assumed the existence of a final zero $M_{z ,(N'+1)}$.
This is specific to to the physical case where there is a percolation path through the $y_2$ bonds but no percolation path through the $y_1$ bonds.
This contribution will be corrected in cases for which such a final zero does not exist.
Using the results in (\ref{bigasy}), in particular on the mean spacing of the zeros between the poles, we may express $P$ as
\begin{align}
P &= D(N) \prod\limits_{k=1}^{N'} \frac{\mu + (f(k) + (1-\bar{\delta}_k) f'(k) )}{\mu + f(k)} \nonumber\\
    &= D(N) \prod\limits_{k=1}^{N'} 1 + \frac{(1-\bar{\delta}_k) f'(k)}{\mu+f(k)} \nonumber.
\end{align}
Now take the logarithm  of both sides and using the approximation $\log(1+x) \approx x$ for small $x$, we have
\begin{align}
\log(P)  &\approx \log(D(N)) + \sum\limits_{k=1}^{N'} \frac{(1-\bar{\delta}_k) f'(k)}{\mu+f(k)}.\label{eqn:sum}
\end{align}
We now approximate the sum in (\ref{eqn:sum}) by an integral, so that
\begin{align}
 \log(P)  &\approx \log(D(N)) + \int\limits_{k=1}^{N'}(1-\bar{\delta}_k) \frac{f'(k)}{\mu+f(k)} \; dk.\nonumber
\end{align}
Making a change of variable from $k$ to $f$, gives
\begin{align}
 \log(P) &\approx  \log(D(N))\nonumber\\
 & +  \int^{M_{p,N'}}_{M_{p,1}}  (1-\bar{\delta}(f)) \frac{df}{\mu+f}.\label{eqn:finally1}
\end{align}
The analysis of this equation depends upon the value of $p$ and we consider separately 
the cases of $p = 1/2$ and $p \ne 1/2$.

\subsection{The asymptotic form of the equations when $p=1/2$}

%%% NDS: begin insert
We  now look at the above equation when $p=1/2$, and first determine the relation between $\eta_k$ and $\delta_k$? 
Let
\begin{align}
\delta W_{p,k} &= W_{p,k+1} - W_{p,k}\nonumber\\
\delta \log W_{p,k} &= \log W_{p,k+1} - \log W_{p,k}.\nonumber
\end{align}
Since
\begin{align}
W_{z,k} &= (1-\delta_k)\delta W_{p,k} + W_{p,k},\nonumber
\end{align}
we have
\begin{align}
\log W_{z,k}  &=  \log \bigl((1-\delta_k)\frac{\delta W_{p,k}}{W_{p,k}} + 1\bigr) + \log W_{p,k}.\nonumber
\end{align}
Comparing with
\begin{align}
\log W_{z,k} &= (1-\eta_k) \delta \log W_{p,k} + \log W_{p,k},\nonumber
\end{align}
and taking Taylor expansions, for $M \rightarrow \infty$, we have
\begin{align}
(1-\eta_k) \sum_{m=1}^M \frac{(-1)^{m+1}}{m!}\biggl(\frac{\delta W_{p,k}}{W_{p,k}}\biggr)^m &= \nonumber\\
\quad \sum_{m=1}^M (1-\delta_k)^m \frac{(-1)^{m+1}}{m!}\biggl(\frac{\delta W_{p,k}}{W_{p,k}}\biggr)^m&.
\end{align}
When $(\delta W_{p,k}/W_{p,k})^2 \ll (\delta W_{p,k}/W_{p,k})$ it follows that
$$\eta_k \approx \delta_k.$$ 
Assuming the log poles have a normal distribution then $\delta \log W_{p,k} \sim {\rm O}(1/N)$. 
For a sufficiently large network, when $p=0.5$, we expect $\delta_k \approx \eta_k$ for most $k$ (the first order Taylor expansion becomes invalid near the tails of the normal distribution, but this contributes relatively little to the summation in (\ref{eqn:sum})).
The results imply that $\bar{\delta}_k$ is very close to being constant at 1/2, so that in (\ref{eqn:finally1}) we have $1 - \bar{\delta} = 1/2.$
%%% NDS: end of insert

We can then integrate the expression for $P$ exactly. This allows sharp estimates of the asymptotic behaviour in this critical case. 
Integrating (\ref{eqn:finally1}) gives
$$\log(P) \approx \log(D(N)) + \frac{1}{2} \log \left( \frac{\mu + M_{p,N'}}{\mu +  M_{p,1}} \right),$$ so that $$P \approx  D(N) \left(\frac{\mu + M_{p,N'}}{\mu + M_{p,1}} \right)^{\frac{1}{2}}.$$

In this {\em critical case} it is equally likely that we will/will not have percolation paths along $y_1$ or $y_2$ bonds at both small and large values of $|\mu|$. 
Accordingly, we must consider four equally likely cases of the distribution of the poles and zeros which could arise in any random realisation of the network.
Thus to obtain the four possible responses of the network we must now consider the contribution of the first zero and also of the 
last zero. 

{\bf Case 1:}  First zero at the origin, last zero at $N'+1$.
This corresponds to there being a  percolation path through the $y_2$ bonds.

To determine this case we  multiply $P$ by $\mu$ to give $Y_1(\mu)$ so that
\begin{equation}
Y_1(\mu) \approx D(N)_1\; \mu \left(\frac{\mu + M_{p,N'}}{\mu + M_{p,1}}\right)^{\frac{1}{2}}.\label{case1}
\end{equation}

{\bf Case 2:}  First zero not at the origin, last zero at $N'+1$.
This corresponds to the existence of percolation paths through $y_1$ bonds and $y_2$ bonds. 

In this case we multiply $P$ by $\mu + M_{z,1}$ to give $|Y(\mu)|$.
We also use the result from the previous section that asymptotically $M_{z,1} \sim M_{p,1}$.
This then gives
\begin{equation}
Y_2(\mu) \approx D(N)_2  \left(\mu + M_{p,N'} \right)^{ \frac{1}{2} } \left( \mu + M_{p,1} \right)^{\frac{1}{2}}.
\label{case2}
\end{equation}

{\bf Case 3:}  First zero at the origin, last zero at $N'$.
Here there are no percolation through either set of bonds.

To determine this case we  multiply $P$ by $\mu$ and divide by $\mu + M_{z,N'}$ to give $Y$. 
Exploiting the fact that asymptotically $M_{p,N'} \sim M_{z,N'}$ we then have
\begin{equation}
Y_3(\mu)  \approx D(N)_3 \; \frac{\mu}{\left( \mu + M_{p,N'} \right)^{\frac{1}{2}}\left(\mu + M_{p,1} \right)^{\frac{1}{2}}} .\label{case3}
\end{equation}

{\bf Case 4:}  First zero not at the origin, last zero at $N'$.
This final case there exists percolation via the $y_1$ bonds  but not through the $y_2$ bonds.

To determine this case we multiply $P$ by $\mu + M_{z,1}$ and divide by $\mu + M_{z,N'}$ to give $Y$. 
Again, exploiting the fact that asymptotically $W_{p,N'} \sim W_{z,N'}$ we have
\begin{equation}
Y_4(\mu) \approx D(N)_4 \; \left( \frac{ \mu + M_{p,1}}{ \mu + M_{p,N'}} \right)^{\frac{1}{2}}.
\label{case4}
\end{equation}

We know, further, from the calculations in the previous section that for all sufficiently large values of $N$ 
$$M_{p,1} \sim  1/N \quad \mbox{and} \quad  \; M_{p,N'} \sim N.$$
Substituting these values into the expression for $Y_1$ gives
\begin{equation}
Y_1(\mu) \approx D(N)_1 \; \mu \left(\frac{\mu + N}{\mu + 1/N}\right)^{\frac{1}{2}}.
\label{case1a}
\end{equation}
The value of the constant $D(N)_1$ can be determined by considering the mid range of each of these expressions in the PLER
The results of the classical Keller duality theory \cite{milton1980bcd} predict that each of the expressions $Y_i$ takes the same form in the range $1/N \ll |\mu| \ll N$ with
\begin{equation}
Y_i(\mu) \approx \sqrt{y_1 y_2}, \quad i=1,2,3,4.
\label{emerge}
\end{equation}
In the case of $Y_1$ we see that the mid-range form of the expression (\ref{case1a}) is given by 
$Y_1 = D_1  \sqrt {N \mu} = \sqrt{N} \sqrt{y_2}/\sqrt{y_1}.$
This then implies that $D_1 = y_1/\sqrt{N}$ so that
\begin{equation}
Y_1(\mu)  \approx \frac{y_1\;  \mu }{\sqrt{N}}  \left( \frac{\mu + N}{\mu + 1/N} \right)^{\frac{1}{2}}.
\label{case1b}
\end{equation}
Very similar arguments lead to the following expressions in the other three cases:
\begin{align}
Y_2(\mu)  &\approx \frac{y_1}{\sqrt{N}} \; (N+\mu)^{ \frac{1}{2} }   \left( 1/N+\mu \right)^{\frac{1}{2}},
\label{case2b1}\\
%\end{align}
%\begin{align}
Y_3(\mu) &\approx  \sqrt{N} y_1 \;  \frac{\mu}{\left( N+\mu \right)^{\frac{1}{2}}\left(1/N+\mu \right) ^{\frac{1}{2}}},
\label{case3b1}\\
%\end{align}
%\begin{align}
Y_4(\mu)  &\approx  \sqrt{N} y_1\; \left(\frac{1/N+\mu}{N+\mu}\right)^{\frac{1}{2}}.
\label{case4b1}
\end{align}
The four formul\ae\  above give a very complete asymptotic description of the response of the binary network when $p = 1/2$. 
In particular they allow us to see the transition between the power-law emergent region and the percolation regions and they also describe the form of the expressions in the percolation regions. 
We see a clear transition between the emergent and the percolation regions at 
\begin{equation}
\mu_1= 1/N \quad \mbox{and} \quad \mu_{2} = N.
\label{limitfreq}
\end{equation}
Hence, the number of components in the system for $p = 1/2$ has a strong influence on the {\em boundaries} of the emergent region and also on the {\em percolation response}. 
However the emergent behaviour itself is {\em independent of $N$}. 
Observe that these frequencies correspond directly to the limiting pole and zero values. 
This gives a partial answer to the question, {\em how large does $N$} have to be to see an emergent response from the network.
The answer is that $N$ has to be sufficiently large so that $1/N$ and $N$ are widely separated frequencies.

The behaviour in the percolation regions in then given by the following:
\begin{align}
Y_1(|\mu| \ll 1)  &\approx y_2 \sqrt{N},   \quad &Y_1(|\mu| \gg 1) &\approx \frac{y_2}{\sqrt{N}},
\label{case1bp}\\
%\end{equation}
%\begin{equation}
Y_2(|\mu| \ll 1)  &\approx \frac{y_1}{ \sqrt{N}}, \quad &Y_2(|\mu| \gg 1) &\approx \frac{y_2}{\sqrt{N}},
\label{case2bp}\\
%\end{equation}
%\begin{equation}
Y_3(|\mu| \ll 1)  &\approx  y_2 \sqrt{N},  \quad &Y_3(|\mu| \gg 1)  &\approx  y_1 \sqrt{N},
\label{case3bp}\\
%\end{equation}
%\begin{equation}
Y_4(|\mu| \ll 1)  &\approx  \frac{y_1}{ \sqrt{N}},  \quad &Y_4(|\mu| \gg 1)  &\approx y_1 \sqrt{N}.
\label{case4bp}
\end{align}
We note that these percolation limits, with the strong dependence upon $\sqrt{N}$ are exactly as observed in Section \ref{sec:models}. 

\subsection{The network response when $p \ne 1/2$} 

This case differs from the case of $p = 1/2$ in a number of ways and the spectral analysis is both harder and less complete. 
Firstly, rather than getting four different responses we expect to see only two. 
When $p > 1/2$ then there will (with probability one) always be conducting capacitative percolation paths for large values of $\omega$ and for small values of $\omega$ we will not get any resistive percolation paths. 
This corresponds to the case of a first zero on the origin and a final zero at $N'+1$. 
Similarly, if $p < 1/2$ then we will get (with probability one) a response with no capacitative percolation path at high frequencies and resistive percolation paths at low frequencies, which corresponds to a first zero away from the origin and no final zero at $N'$.
Hence, we need only consider {\bf Case 1} and {\bf Case 4} respectively. 
Secondly the values for the conductance at high and low frequencies are asymptotically independent of (sufficiently large) $N$. Furthermore
the formula for $P$ in (\ref{eqn:finally1})  involves a quadrature involving $1 - \bar{\delta}$ which cannot be obtained in closed form. As a consequence
we shall adopt a different approach for this case by combining the spectral calculation with that of averaging.

\section{Effective medium (averaging) calculations}\label{sec:ema}

The Effective Medium Approximation (EMA) formula derived by an averaging method \cite{Kirk},  gives an approximation to the conductance of the network,
and is derived by regarding the random distribution of the random resistors and capacitors as a series of perturbations
of a uniform field identical conductors. The conductance of the effective medium is chosen to minimise the first moment
of the resulting perturbation matrix. It assumes an infinitely large number of conductances and hence corresponds to taking
$N \to \infty$ in the previous analyses. Whilst accurate for $p$ not too close to 1/2 it has limitations for $p$ close to 1/2, in that
whilst it predicts a transition from emergent to percolation type behaviour, the form of this transition is not
quite correct as $p \to 1/2$. Thus the EMA calculations are complimentary to those derived using spectral methods in the previous section. 
In this section we will review the EMA result, and show that it is consistent with a PLER description of the 
behaviour with a power law which we explicitly derive. We we then make a (somewhat speculative,
but consistent) extension of the EMA calculation to include the effects of finite network size $N$.
We see presently that if 
$N > N^* \equiv |p - 1/2|^{-2}$
then the EMA formula gives a good approximation to the resulting conductance and we will obtain a more general
formula which is effective for all $|p-1/2|$ and $1/\sqrt{N}$ sufficiently small.

\subsection{Infinite networks}

\subsubsection{Overview}

If the conductances are $y_1$ and $y_2$ are in respective proportion $1-p$ and $p$ then, from the `classical' EMA result, \cite{Kirk},
the effective medium conductance $Y$ for a very large $(N \to \infty)$ {\em square two-dimensional lattice} is given by the solution of the
quadratic equation
\begin{equation}
{
(1-p) \left( \frac{Y - y_1}{Y+y_1} \right) + p \left( \frac{Y - y_2}{Y+y_2} \right) =0.
}
\label{aprema1}
\end{equation}
Rearranging this formula we have
$$Y^2 + (1-2p)(y_2 - y_1) Y - y_1 y_2 = 0,$$
so that if
$$\epsilon = (1-2p), \quad \theta = Y/\sqrt{y_1 y_2}, \quad \mu = y_2/y_1,$$
we have
\begin{equation}
{
 \theta - \frac{1}{\theta} + \epsilon \left( \sqrt{\mu} - \frac{1}{\sqrt{\mu}} \right) = 0.
 }
 \label{ema42}
 \end{equation}
Now set
$$\gamma = \log(\theta), \quad \mbox{and} \quad  \nu = \log(\mu)$$
it follows immediately that
\begin{equation}
{
\sinh(\gamma) = -\epsilon \sinh(\nu/2).
}
\label{aprema2}
\end{equation}

\subsubsection{Emergent power laws}

Suppose firstly that $\mu$ is {\em real}, so that we are modelling a R-R network, and that $\mu$ is close to unity, so that $\nu$, and hence $\gamma$
are both not large. Then we may linearise
(\ref{aprema2}) and to leading order  we have
$\gamma = -\epsilon \nu/2.$
Thus in this case $\log(Y/\sqrt{y_1 y_2}) = \epsilon \log(y_1/y_2)/2$, and rearranging this gives the elegant power law identity
\begin{equation}
{
Y = y_1^{(1-p)} y_2^{p}.
}
\label{aprema3}
\end{equation}
This is fully consistent with the duality result (\ref{dual}) that
$$Y(y_1,y_2) Y(y_2,y_2) = y_1^{(1-p)} y_2^p y_1^p y_2^{(1-p)} = y_1 y_2.$$

In the C-R network case, $\mu = i \omega CR$ is pure imaginary. We set $\mu = i \eta$ where $\eta = \omega CR$ is now assumed to be close to unity and take $\beta = \log(\eta)$
to be close to zero. It then follows that $\log(\mu) = i \pi/2 + \beta$ so that
$$\sinh(\gamma) = -\epsilon \sinh( i\pi/4 + \beta/2)  = -\epsilon(i + \beta/2 + {\cal O}(\beta^2) )/\sqrt{2}.$$
If $\beta = 0$ then 
$\gamma = i \theta_0$ where $\sin(\theta_0) = -\epsilon/\sqrt{2}.$
Linearising about this solution we then have, to leading order,
$$\gamma = i \theta_0 - \frac{\epsilon}{2\sqrt{2} \cos(\theta_0)}\beta + {\cal O}(\beta^2) \equiv i \theta_0 + \Lambda \log(\eta) + {\cal O}(\beta^2).$$
Thus, to leading order
$$|Y| = \sqrt{\omega C/R}\  |\exp(\gamma)| = \sqrt{\omega C/R} \; (\omega CR)^{\Lambda} \equiv K \omega^{\alpha}.$$
This gives precisely the power law observed in the results presented in Section \ref{sec:models} in a frequency range centred around the region for which
 $\omega CR = {\cal O}(1)$ and
 \begin{equation}
 {
 \alpha(p) = \frac{1}{2} + \Lambda = \frac{1}{2} - \frac{\epsilon}{2 \sqrt{2} \sqrt{1 - \epsilon^2/2}}.
 }
 \label{snow10}
\end{equation}

Note that if $p = 1/2$, so that $\epsilon = 0$, then we see power law behaviour with exponent $1/2$ in all cases, over the
entire range. This contrasts with the results of the last section, in which we see percolation type behaviour
for (say) $\omega > N$, but this also emphasises the fact that the EMA calculation only applies for $N = \infty$ in this case. 
Note further that $\alpha(0)=0,\alpha(1/2)=1/2,\alpha(1)=1$ and that $\alpha(p) \approx p$ for all $0 \le p \le 1$.

\subsubsection{Percolation  behaviour}

It is well known that the EMA approximation for $p \ne 1/2$ exhibits percolation behaviour which we summarise here.
If we consider the case of large positive  $\nu$ then the asymptotic form of the solution
depends upon the sign of $\epsilon$. If $\epsilon > 0 $, which corresponds to $p < 1/2$ then for large $\alpha$ the equation
(\ref{aprema2}) reduces to $e^{-\gamma} = \epsilon e^{\nu/2}$
so that we have the percolation behaviour given by: 
\begin{equation}
{
 1/\theta = \epsilon \sqrt{\mu}, \quad \quad Y = y_1/\epsilon
}
\label{aprema5}
\end{equation}
In contrast, if $\epsilon < 0$, ($p > 1/2$) then for large $\nu$ the equation (\ref{aprema2}) simplifies to:
$e^{ \gamma} = (-\epsilon) e^{\nu/2}$
so that we have the percolation  behaviour given by:
\begin{equation}
{
\theta = -\epsilon \sqrt{\mu}, \quad \quad Y = -\epsilon y_2.
}
\label{aprema6}
\end{equation}
Similar results for $\nu$ large and negative follow from duality arguments. 
The (frequency) limits of the emergent  region can be estimated by finding when the power law behaviour of (\ref{aprema3}) overlaps with the percolation
type behaviour.  This leads to the following estimates for the values $\mu_1 < \mu < \mu_2$ over
which we expect to see power-law emergent behaviour
\begin{align}
\epsilon > 0:&\quad  \mu_1 = 1/\mu_2  \sim \epsilon^{1/p},\nonumber\\
\epsilon < 0:&\quad  \mu_1 = 1/\mu_2  \sim (-\epsilon)^{1/(1-p)}.
\label{aprema8}
\end{align}

{\bf Important Note}  These results predict that as $\epsilon \to 0$ the percolation amplitude scale as $|\epsilon|^{\pm 1}$. This is not
quite what is observed in practice. In contrast, empirical calculations described in (for example) \cite{jonckheere1998drb} , imply instead a scaling
law of the from $|\epsilon|^{\pm 1.3}.$ Thus the EMA is not fully accurate in this limit. 

\subsection{Large, but finite, networks}

We now give a more speculative calculation which attempts to combine the EMA estimate with finite size effects and the spectral calculations of the previous 
section, for Case 1 and Case 4. Our starting point is the spectrally derived formula $Y_1$ (\ref{case1b}) which has percolation limits proportional to $y_2$ when $\mu$ is large.
Casting $Y_1$ in terms of $y_1$ and $y_2$ we have
\begin{align*}
Y_1^2  &= \frac{\mu^2 y_1^2}{N} \; \frac{(\mu + N)}{(\mu + 1/N)}\\
	&= \mu y_1^2 \; \frac{(1 + \mu/N)}{(1 + 1/N\mu)}\\
 	&= y_1 y_2 \; \frac{(1 + \mu/N)}{(1 + 1/N\mu)}.
\end{align*}
It follows that
$$(1 + 1/N\mu)Y^2 - y_1 y_2 (1 + \mu/N) = 0.$$
If, as above, we set $\theta = Y/\sqrt{y_1 y_2}$ we have, after some manipulation, that this formula has the symmetric form
\begin{equation}
{
\theta - \frac{1}{\theta}= \frac{1}{N} \left( \frac{\mu}{\theta} - \frac{\theta}{\mu} \right).
}
\label{rainy1}
\end{equation}
Similarly, the spectrally derived formula $Y_4$ in (\ref{case4b1}), which has percolation limits proportional to $y_1$ for $\mu$ large, takes the symmetric form
\begin{equation}
{
\theta - \frac{1}{\theta} = \frac{1}{N} \left( \frac{1}{\mu \theta} - \mu \theta \right).
}
\label{rainy2}
\end{equation}
These expressions are both very similar in form to the  result  (\ref{ema42}) of the EMA calculation. We conjecture that a more general expression can be obtained 
by combining them into the following  two
fomul\ae\ which agree with each in the limits of $\epsilon = 0$ and $N = \infty$ and which include both the effects of component 
proportion and network size and which respectively have percolation limits proportional to $y_1$ and $y_2$:
\begin{equation}
{
\theta - \frac{1}{\theta} + \epsilon \left( \sqrt{\mu} - \frac{1}{\sqrt{\mu} } \right) = \frac{1}{N} \left( \frac{1}{\mu \theta} - \mu \theta  \right)
}
\label{rainy1a}
\end{equation}
and
\begin{equation}
{
\theta - \frac{1}{\theta} + \epsilon \left( \sqrt{\mu} - \frac{1}{\sqrt{\mu} } \right) = \frac{1}{N} \left( \frac{\mu}{\theta} - \frac{\theta}{\mu} \right).
}
\label{rainy2a}
\end{equation}

We observe that each of (\ref{rainy1a}) and (\ref{rainy2a}) is {\em self-dual} under the map $\mu \to 1/\mu$, $\theta \to 1/\theta$.
Similarly the symmetry $\mu \to 1/\mu$, $\epsilon \to -\epsilon$ maps (\ref{rainy1a}) to (\ref{rainy2a}) and vice versa. We now
proceed to show that (\ref{rainy1a},\ref{rainy2a}) have the correct asymptotic form of solution and give
numerical evidence for their validity in Section \ref{sec:comparison}. We firstly consider the percolation limits of (\ref{rainy1a}) and (\ref{rainy2a}). Motivated by the analysis in the previous subsection we consider solutions
of the form $\theta = \beta \sqrt{\mu}$ so that $Y = \beta y_2$, and $\theta = \beta/\sqrt{\mu}$, so that $Y = \beta y_1$, in the two cases of $\mu$ large and $\mu$ small. If $\mu$
is large and $\epsilon > 0$ then (\ref{rainy1a}) has a solution with percolation limit proportional to, and in phase with, $y_1$, so that $\theta = \beta/\sqrt{\mu}$ if $\beta$ satisfies
the quadratic equation
\begin{equation}
{
\frac{\beta^2}{N} + \epsilon \beta - 1 = 0.
}
\label{rainy3}
\end{equation}
If $\mu$ is small then we have the reciprocal solution given by the map $\beta \to 1/\beta$. 
Note that if $\theta = \beta/\sqrt{\mu}$ then
$|Y| = \beta |y_1|$ and hence the dynamic range is given by
\begin{equation}
{
\hat{Y} = \beta^2 \quad \mbox{where} \quad \frac{\beta^2}{N} + \epsilon \beta = 1.
}
\label{rainy4}
\end{equation}
It is immediate that $\beta$ is given by
\begin{equation}
\beta = \frac{N}{2} \left( -\epsilon  +  \sqrt{ \epsilon^2  + 4/N} \right),
\label{betaeqn}
\end{equation}
where the positive sign for the square root term is taken to ensure that $\beta > 0$ so that the response is in phase with $y_1$. This expression takes two different
forms depending on whether {\em (i)} $N \ll \epsilon^{-2}$ or {\em (ii)} $N \gg \epsilon^{-2}$.   In the first case  the network behaves in a similar way to one with $p = 1/2$
and we have $\beta \sim \sqrt{N}$. In the second case we have behaviour similar to $N = \infty$ with $\beta \sim 1/\epsilon.$ This is in exact correspondence with the
calculations of the dynamic range reported in Section \ref{sec:models}.
Similarly, if $\mu$ is large and $\epsilon < 0$ then the equation (\ref{rainy2a}) has a solution with percolation limit proportional to ,and in phase with,  $y_2$, so that $\theta = \beta \sqrt{\mu}$, if $\beta$ satisfies
 the quadratic equation

\begin{equation}
{
\beta^2 + \epsilon \beta - \frac{1}{N} = 0.
}
\label{rainy5}
\end{equation}

\section{Comparison of the asymptotic and numerical results}\label{sec:comparison}

We now give two sets of calculations for finite networks.  The first tests the validity of the spectral calculation in Section \ref{sec:asymp} for the case of $p = 1/2$. The second the validity of the amalgamated spectral
and averaging based calculation in Section \ref{sec:ema}.

\subsection{Spectral based calculations for  $p = 1/2$ } 

\begin{figure}
\centering
\includegraphics[width=0.8\linewidth]{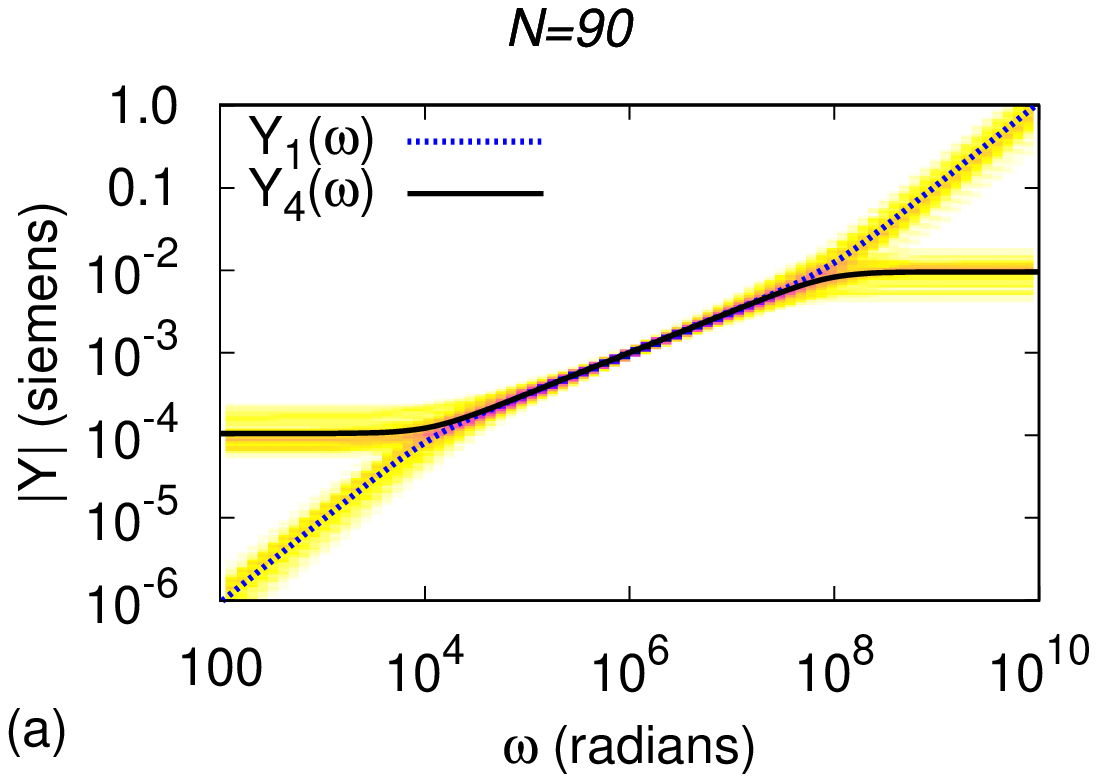}
\includegraphics[width=0.8\linewidth]{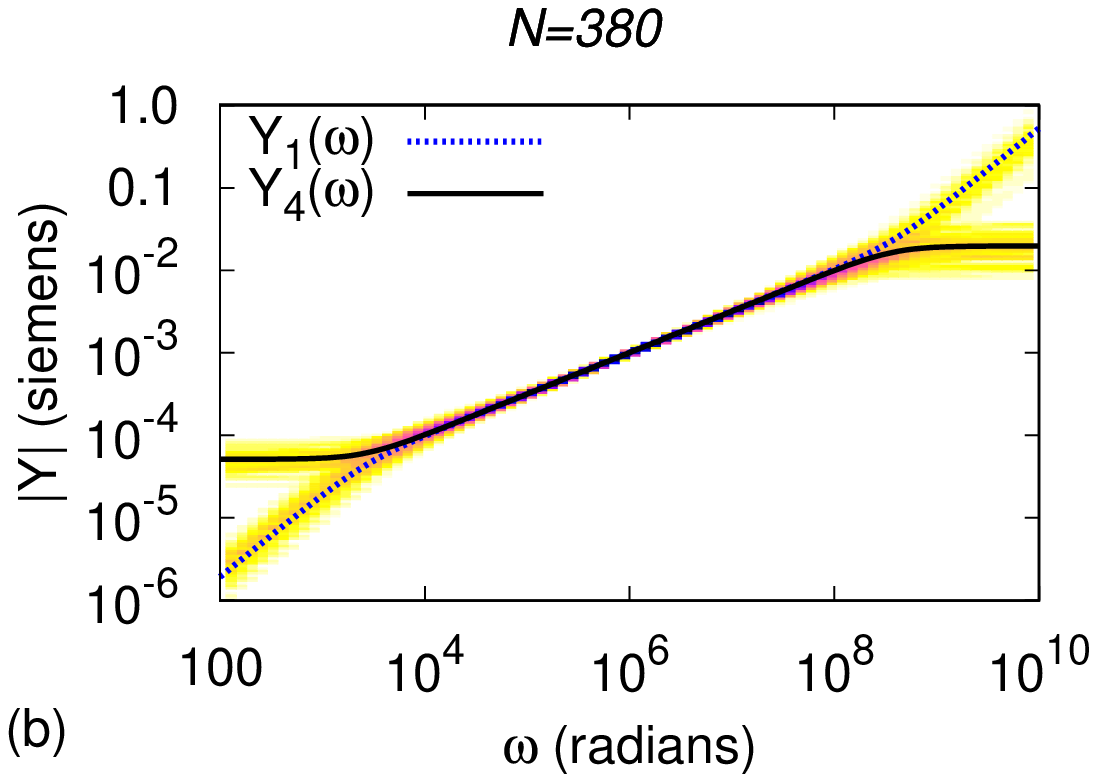}
\includegraphics[width=0.8\linewidth]{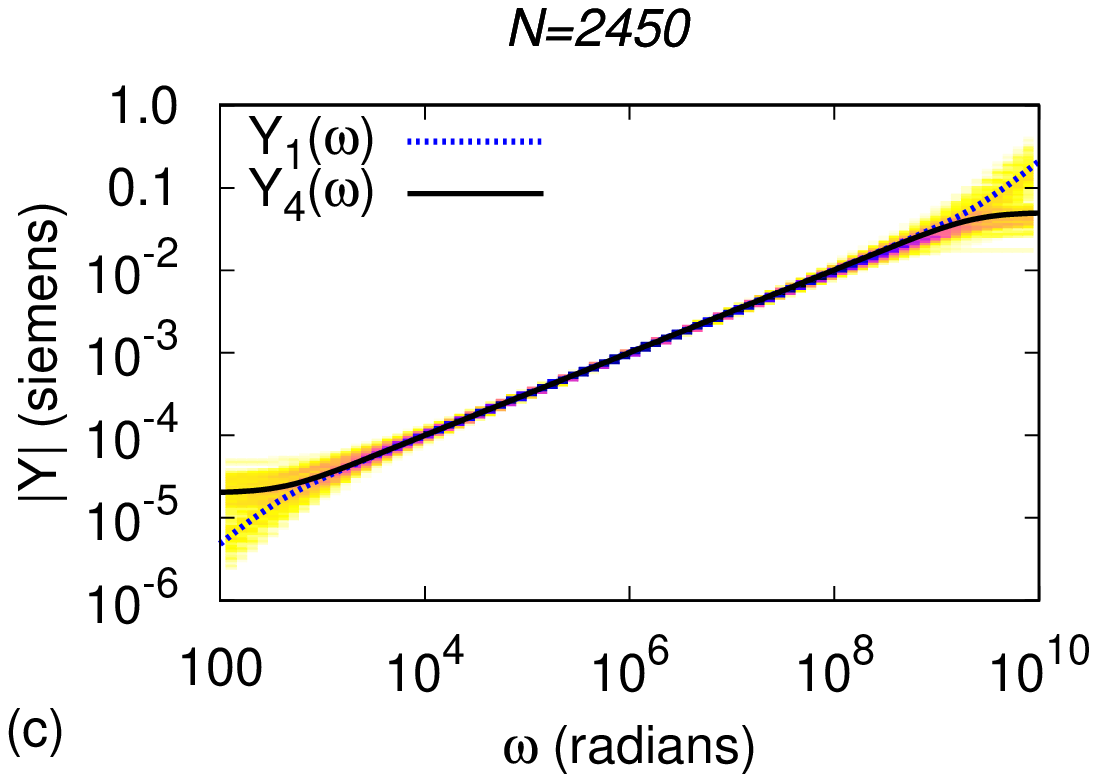}
\includegraphics[width=0.8\linewidth]{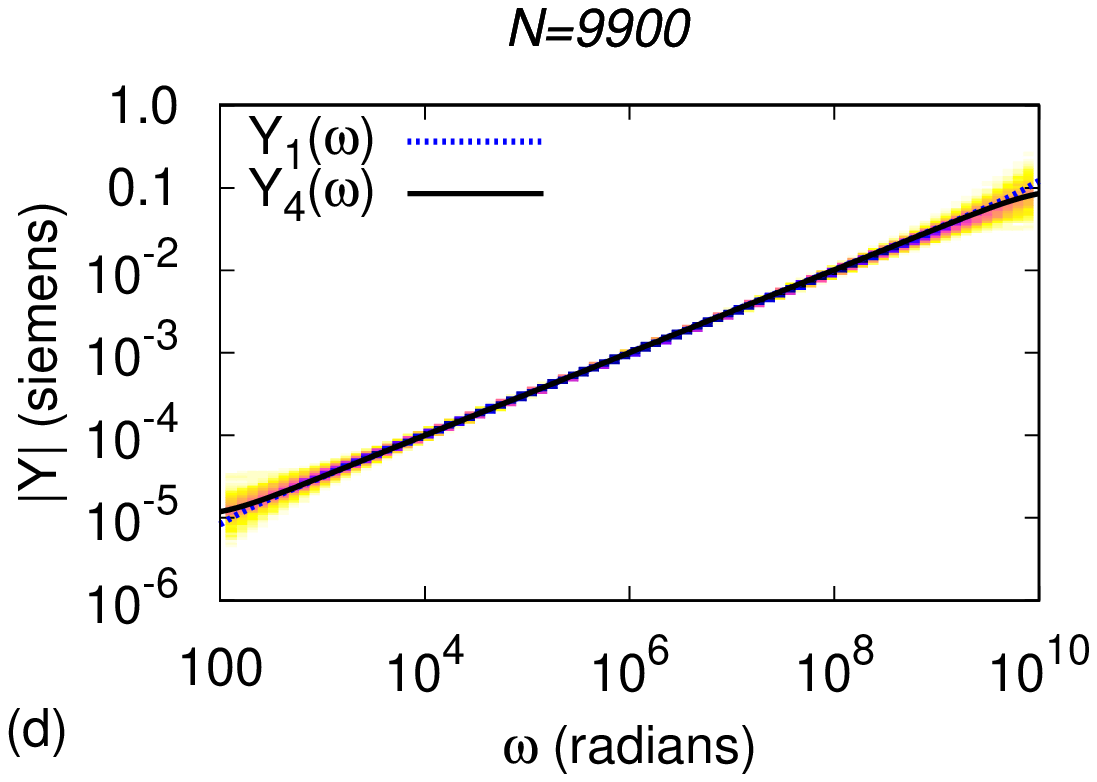}
\caption{(color online) Comparison of the asymptotic formul\ae\  obtained from the spectral derivation
with the numerical computations for the $C-R$ network over many runs, with $p=1/2$ and network sizes sizes $S=10,20,50,100$, $N=S(S-1)$.}\label{fig:compareRC}
\end{figure}

We firstly consider the case of $p=1/2$ ($\epsilon = 0$) for the C-R network with complex  $\mu = i \omega C R$. We compare the absolute values of the four asymptotic formul\ae\  (\ref{case1b},\ref{case2b1},\ref{case3b1},\ref{case4b1}) 
obtained by using the spectral method with the numerical calculations of the absolute network conductance $|Y|$ with $C = 1nF$ and $R = 1k\Omega$ as a function of $\omega$ for four different configurations of the system, with 
different percolation paths for low and high frequencies. 
The results of this comparison are shown in Figure \ref{fig:compareRC} in which we plot the numerical calculations together with the asymptotic formul\ae\  for a range of values of $N$ given by $N=S(S-1)$ with $S=10,20,50,100$.  
We can see from this that  the predictions of  the asymptotic formul\ae\  (\ref{case1b},\ref{case2b1},\ref{case3b1},\ref{case4b1}) fit perfectly with the results of the numerical computations over all of the values of $N$ considered. 
Indeed they agree both in the (square-root) power law emergent region and in the four possible percolation regions. 
The results and the asymptotic formul\ae\ clearly demonstrate the effect of the network size in these cases.
 
As a separate calculation, we look at the results obtained for the R-R network with conductances $1/R$ and $\mu/R$ with real $\mu$ and with $R$ as above.
Again we compare the predictions of the asymptotic formul\ae\  (\ref{case1b},\ref{case2b1},\ref{case3b1},\ref{case4b1}) with the numerical computations of $|Y|$ in this case.
\begin{figure}
\centering
\includegraphics[width=0.8\linewidth]{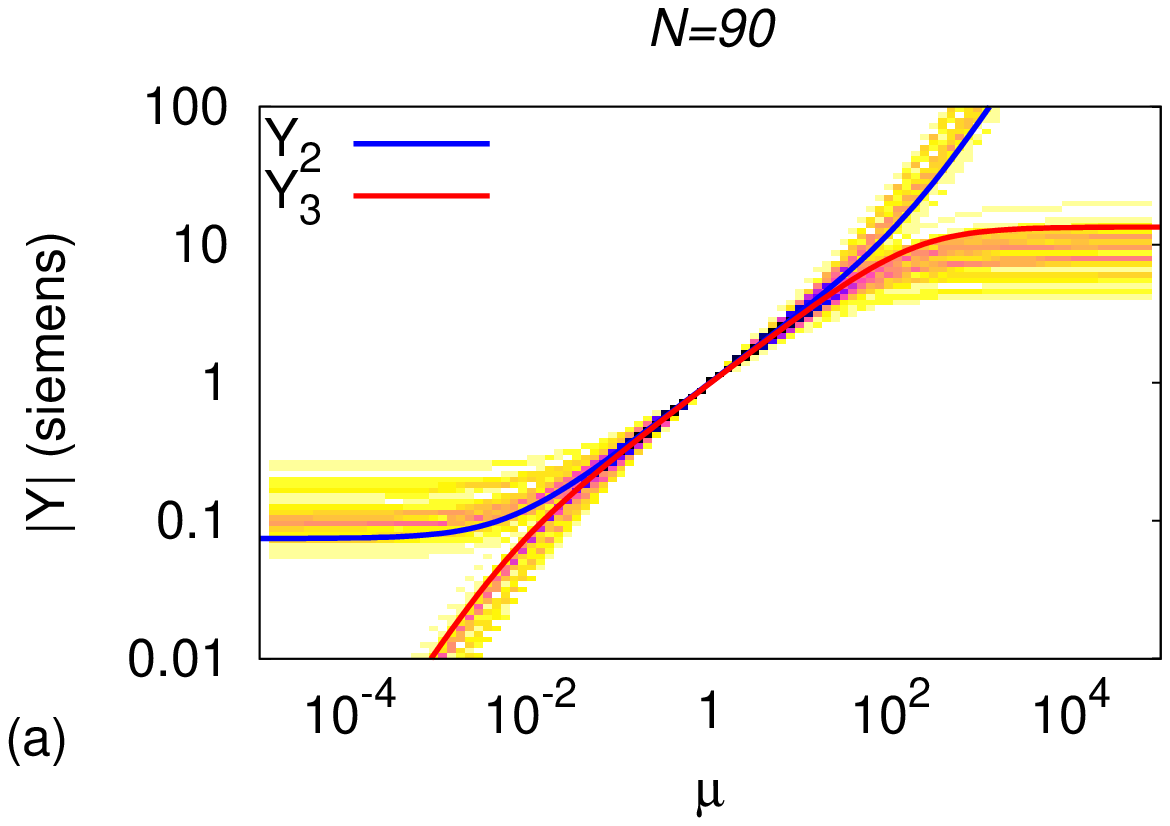}
\includegraphics[width=0.8\linewidth]{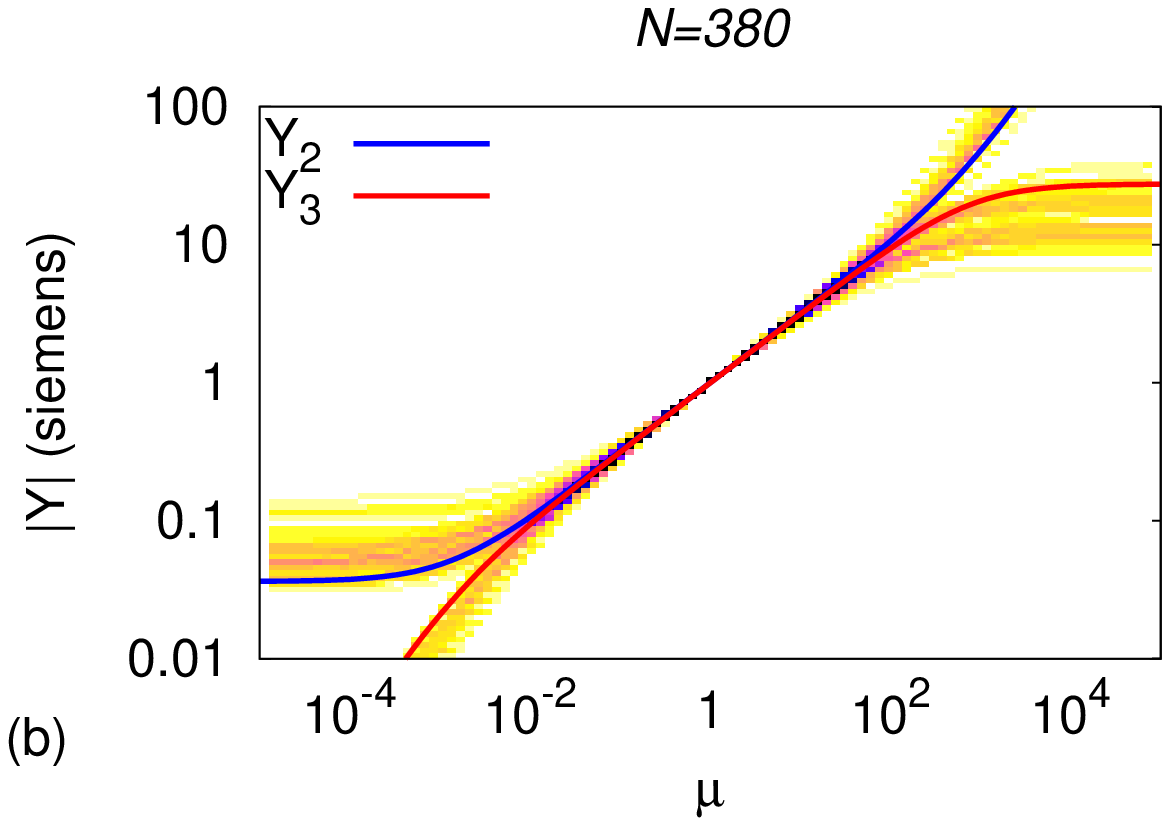}
\includegraphics[width=0.8\linewidth]{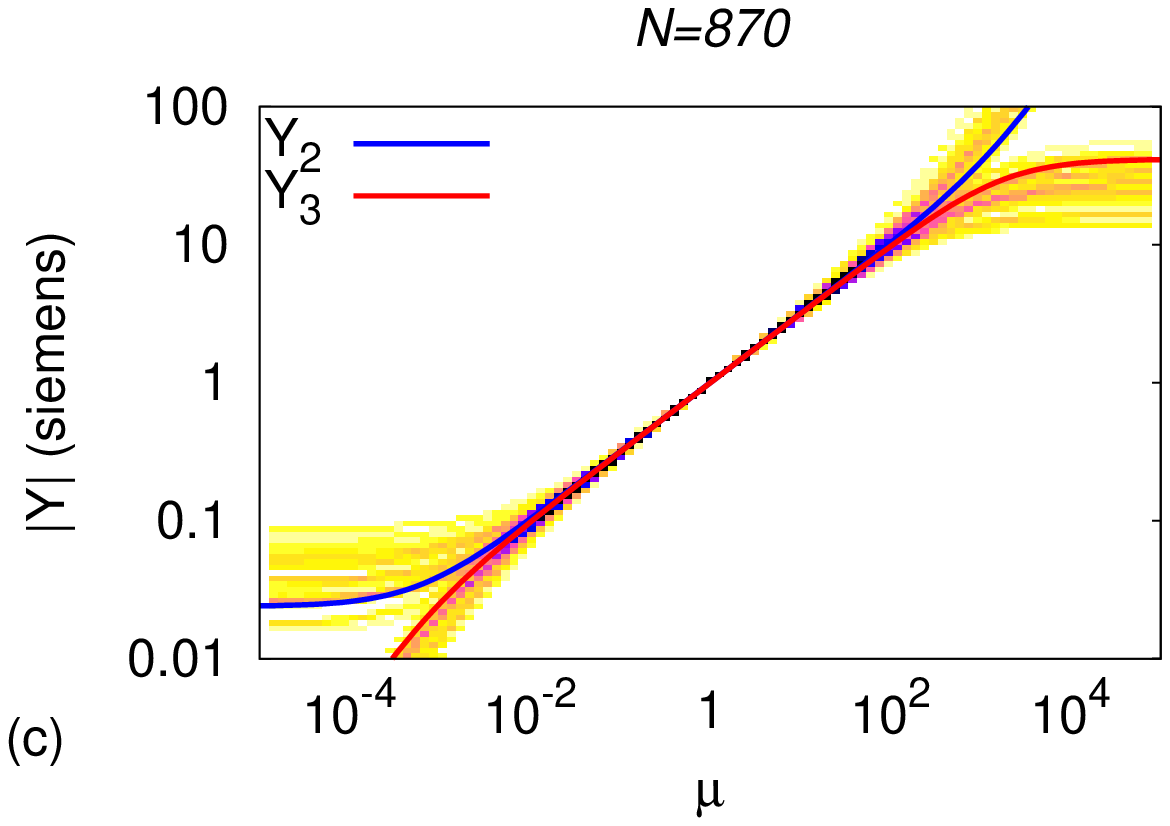}
\includegraphics[width=0.8\linewidth]{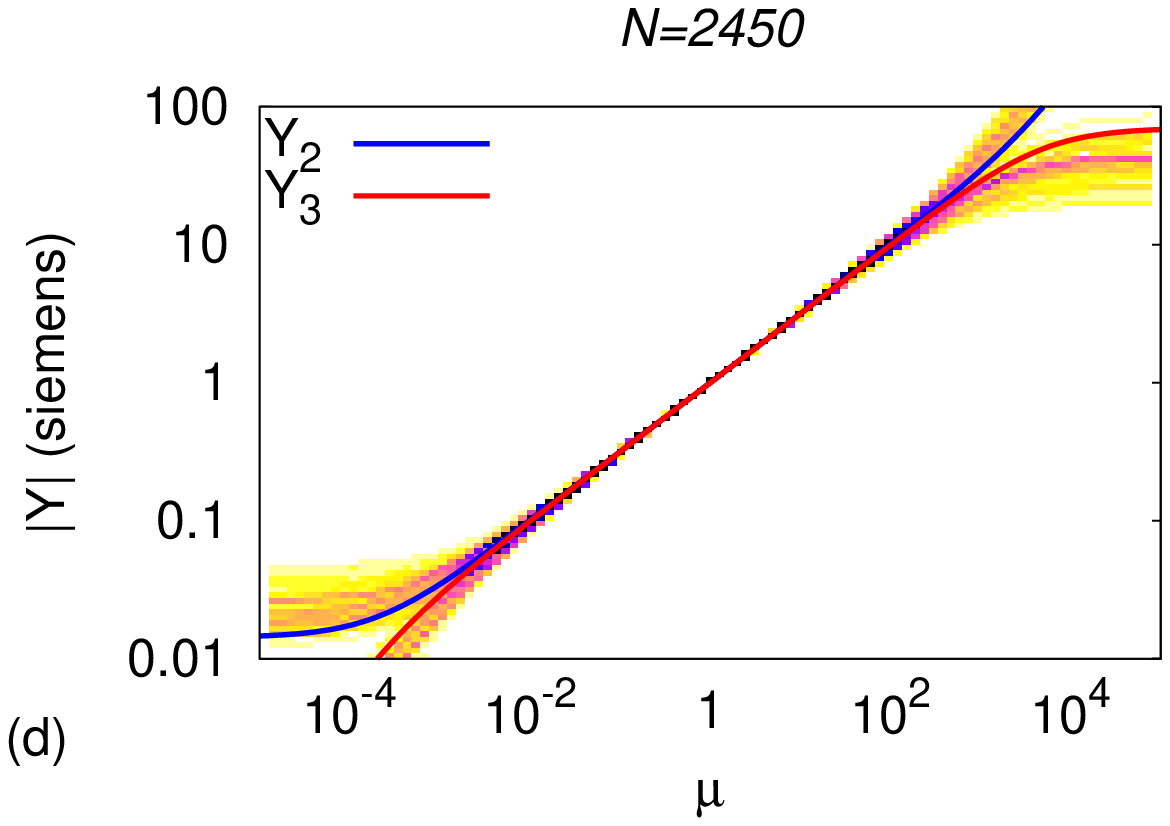}
\caption{(color online) Comparison of the asymptotic formul\ae\  with the numerical computations for the $R-R$ network over many runs, with $p=1/2$ and network sizes sizes $S=10,20,50,100$, $N=S(S-1)$.}\label{fig:compareRR}
\end{figure}
Again we see an excellent agreement in all cases, as shown in Figure \ref{fig:compareRR}.

\subsection{Combined averaging and spectral based calculations for  general $p$}\label{sec:issues}

We now consider the responses described by the pair of quadratic equations  (\ref{rainy1a},\ref{rainy2a}) for $Y$ obtained
by combining the averaging and spectral calculations. We compare these with numerical results for the C-R networks described in the previous 
sub-section.  In each case of $p$ we take the equation for which the corresponding solution in the percolation regime is physically correct.

\begin{figure}[h!]
\centering
\includegraphics[width=0.8\linewidth]{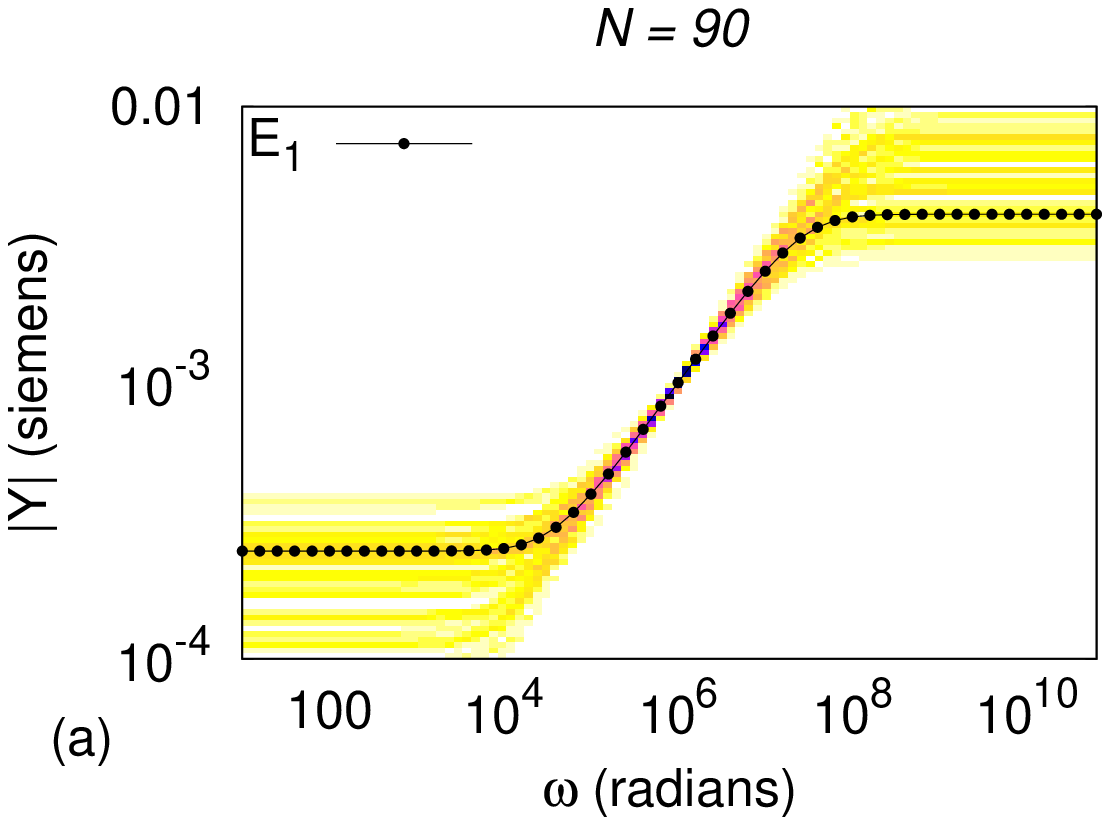}
\includegraphics[width=0.8\linewidth]{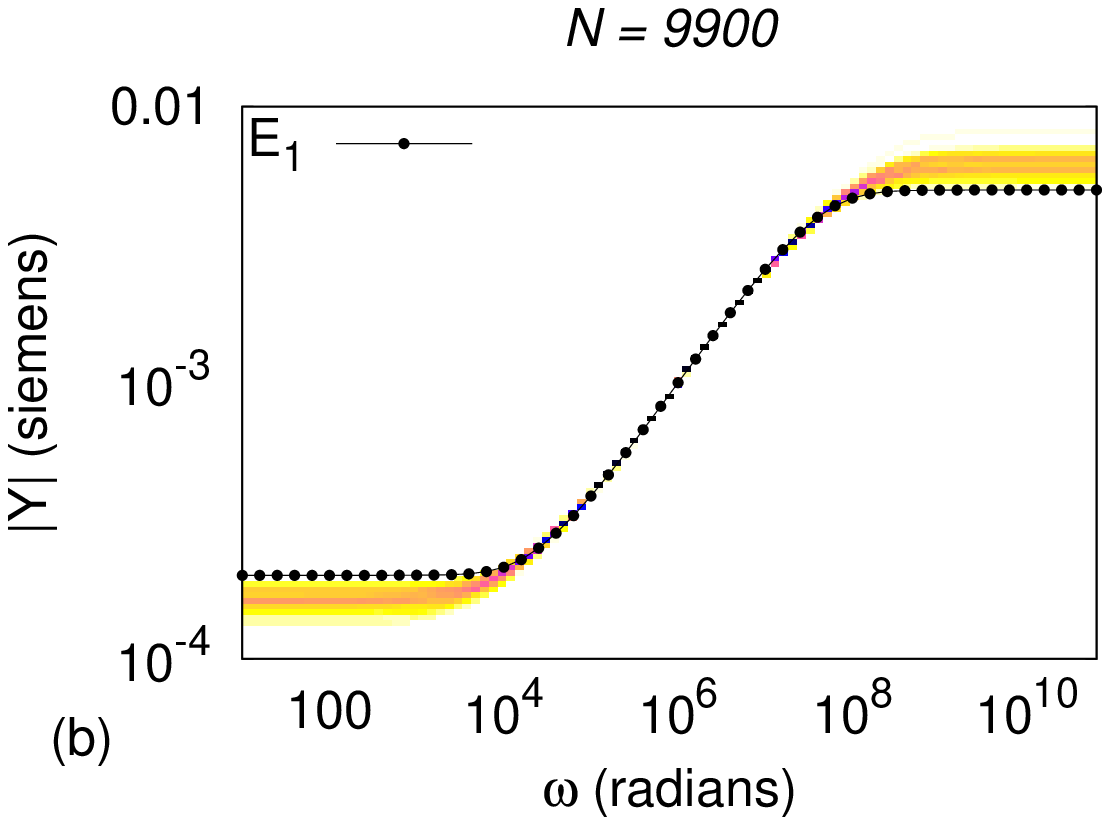}
\caption{(color online) Comparison of the numerically obtained solutions of the quadratic equation (\ref{rainy1a}) with 
the numerical computations for many realisations of the the $C-R$ network, with $p=0.4$ and network sizes sizes $S=10,100$, $N=S(S-1)$.}\label{fig:compareRC69}
\end{figure}

As a first computation we take $p = 0.4$, so that $\epsilon = 0.2 > 0$, and consider a C-R network with the same values of $C,R$ and taking $S$ so that $N = 90,9900 $.
For an infinitely large network we expect to see resitive type percolation behaviour (with probability one) for both small and large frequencies.
We compute $|Y(\omega)|$ from (\ref{rainy1a}) and compare these values with the results of computations of $|Y(\omega)|$ from a large realisations of the network in
Figure \ref{fig:compareRC69}.
The results from this computation are  interesting. We see in the case of $N = 90$, that there is quite
a large statistical range in the calculations in this case. In the case of $N = 9900$ 
 the statistical range
is much smaller. The results of the predictions of $|Y|$ from (\ref{rainy1a}) closely
follow the mid-range of the computations for $N = 90$. In the case of $N = 9900$ there continues to be good agreement,
however, as expected from the EMA results, the maximum and minimum values of $|Y|$ are slightly underestimated by (\ref{rainy1a}).
In both cases the PLER is very well approximated by the computations from (\ref{rainy1a}).
The results for computations of the R-R network are very similar and we do not include them here.

\begin{figure}[h!]
\centering
\includegraphics[width=0.8\linewidth]{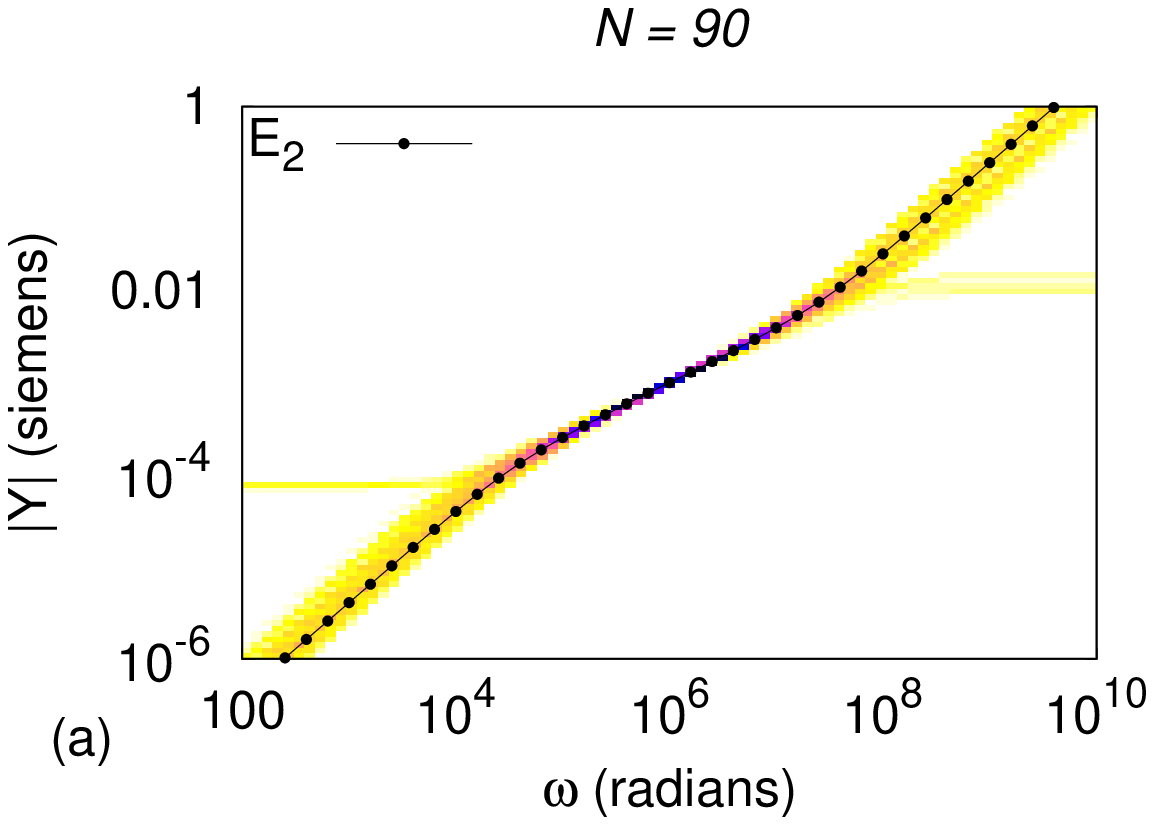}
\includegraphics[width=0.8\linewidth]{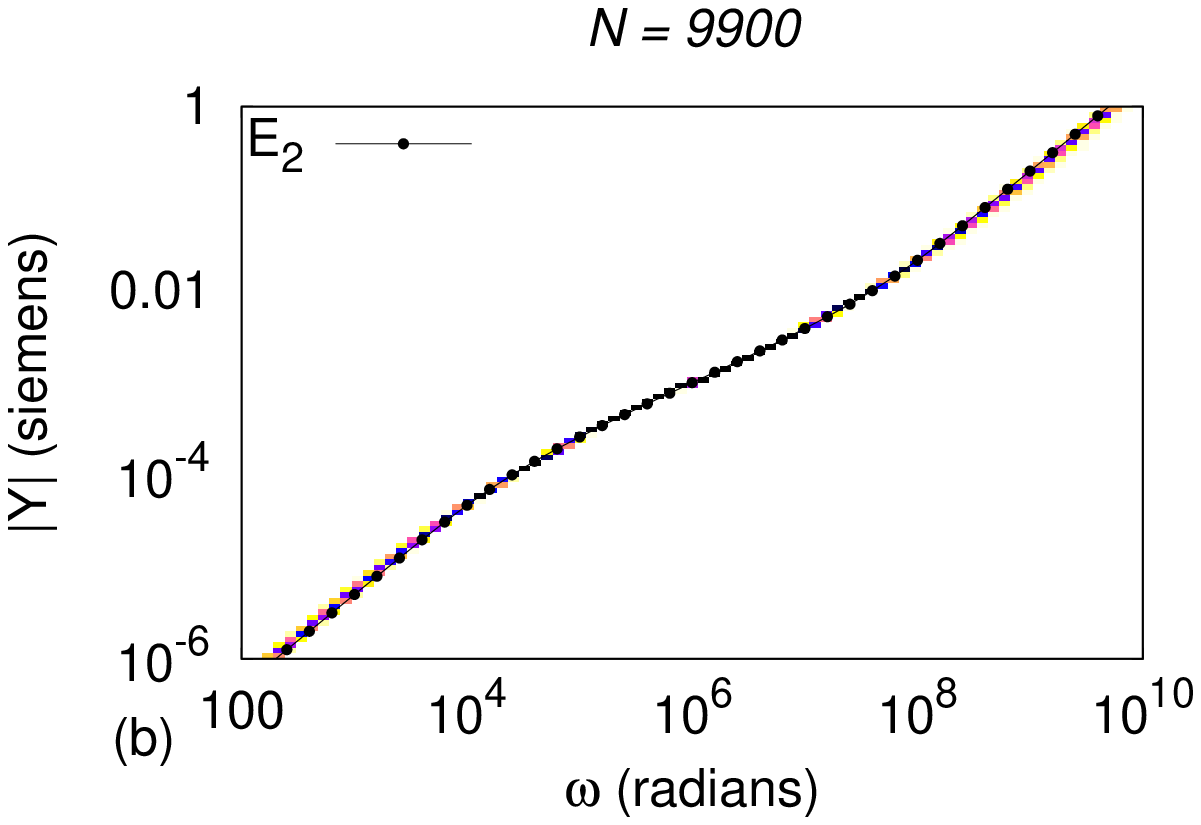}
\caption{(color online) Comparison of the numerically obtained solutions of the quadratic equation (\ref{rainy2a}) with 
the numerical computations of $|Y(\omega)|$ for many realisations of the  $C-R$ network
with $p=0.6$ and network  sizes $S=10,100$, $N=S(S-1)$.}\label{fig:compareRCS69}
\end{figure}

As a second computation we take $p = 0.6$, so that $\epsilon = -0.2 < 0$.
 For an infinitely large C-R network we expect to see reactive type percolation behaviour (with probability one) for both small and large frequencies. We present the results of computing $|Y(\omega)|$ from the quadratic
equation (\ref{rainy2a}) compared with computations from a number of realisations of the network when $N=90, 9900$, in Figure \ref{fig:compareRCS69}.
In this computation we again see a greater statistical range when $N = 90$ than when $N =9900$. Indeed
in the case of $N = 90$ a small number of the realisations show resistive percolation behaviour rather than reactive. This is
not seen in the calculations for $N = 9900$. In both cases the results of the calculations from (\ref{rainy2a}) closely match
the computations over the whole range.

\section{Discussion}\label{sec:concs}

We have attempted to answer questions surrounding emergent behaviour, determining what causes it, finding the range of parameters over which it applies and addressing the question of which aspects of a complex system
influence the emergent behaviour.
Using  large binary networks we have shown how power law emergence can be directly related to the statistical regularity of the spectrum of the matrices associated with the network and hence can be studied by combing
spectral and averaging methods.
In particular we have studied the effects of network size, and the variation from criticality on the observed power law behaviour of these systems.
We have shown how the response of the networks depends strongly upon $p$ and less strongly on the network size $N$, except at $p=1/2$ exactly, where the \emph{dynamic range} has been found to scale in direct proportion to $N$.
When $p = 1/2$ we analysed how the network response is described in terms of poles and zeros of the conductance and can be determined from distribution of these values,
making use of numerically observed statistical patterns of these. 
This has revealed four asymptotic formul\ae, corresponding to the four qualitatively different \emph{emergent responses} that can arise when $p = 1/2$ and these show very precisely
the effects of the (finite) network size $N$.
The case of $p=1/2$ is very complete asymptotically and shows particularly good agreement with the numerical computations, which is remarkable given the number of approximations made.
An important open question is to now rigorously establish the observed statistical results of the spectrum in this case, for example to show rigorously that $\mu_{p,N'} \sim N$.
When $p \ne 1/2$ the analysis is less complete. It is interesting, however, that the results of the averaging based EMA calculation can be combined with those of the spectral computation in a consistent
manner to the case of finite $N$, leading to predictions (\ref{rainy1a},\ref{rainy2a}), of the conductance and its dynamic range which is in good qualitative agreement with what
is observed. However a limitation of this analysis remains the lack of precision of the estimation of the power law scaling of the magnitude of the
percolation response as $p \to 1/2.$
We conclude that combining both the spectral based and the averaging based methods lead to useful asymptotic formul\ae\ with excellent numerical support, and establishing these more
rigorously is an interesting area of further study.

\bibliography{ref}
\bibliographystyle{unsrt}

\end{document}